\documentclass[10pt,twoside]{article}

\usepackage[
    a4paper,
    top=21mm, bottom=20mm,
    left=14mm, right=14mm,
    columnsep=5mm
]{geometry}

\usepackage[T1]{fontenc}
\usepackage[scaled]{helvet}

\usepackage{microtype}

\usepackage{multicol}           %

\usepackage{amsmath,amssymb}

\usepackage{graphicx}
\usepackage{tikz}
\newcommand{\DMPixelSizeNm}{65}               %
\newcommand{\TiltPixelSizeNm}{97}             %

\newcommand{\DMImageBarUm}{3}
\newcommand{\DMZoomBarUm}{1}
\newcommand{\DMPSFBarUm}{0.5}
\newcommand{\TiltImageBarUm}{10}
\newcommand{\TiltZoomBarUm}{2}
\newcommand{\TiltPSFBarUm}{1}
\newcommand{\ActinOverviewBarUm}{20}
\newcommand{\ActinDetailBarUm}{1}

\DeclareRobustCommand{\ScaleLength}[1]{#1\,\mu\mathrm{m}}

\newcommand{\FigureWithScaleBars}[5][\linewidth]{%
    \begin{tikzpicture}
        \node[anchor=south west,inner sep=0,outer sep=0] (scaleimage)
        at (0,0) {\includegraphics[width=#1]{#2}};
        \begin{scope}[x={(scaleimage.south east)},y={(scaleimage.north west)}]
            \def\ScaleCanvasWidth{#3}%
            \def\ScaleCanvasHeight{#4}%
            #5
        \end{scope}
    \end{tikzpicture}%
}
\tikzset{panellabel/.style={anchor=north west,inner sep=1.5pt,font=\normalsize\rmfamily\bfseries}}
\newcommand{\PanelLabel}[3]{%
\node[panellabel] at ({#1/\ScaleCanvasWidth},{1-#2/\ScaleCanvasHeight}) {(#3)};}

\newcommand{\PanelLabelLeft}[3]{%
\node[panellabel,anchor=north east] at (0,{1-#2/\ScaleCanvasHeight}) {(#3)};}
\newcommand{\ScaleBarLabelShift}{0pt}
\newcommand{\ScaleBarLabel}[3]{%
    \node[anchor=south,inner sep=0.7pt,yshift=1.2pt,xshift=\ScaleBarLabelShift,
    font=\fontsize{6}{7}\selectfont,text=white,fill=black,fill opacity=.55,text opacity=1]
    at ({(#1)/\ScaleCanvasWidth},{1-(#2)/\ScaleCanvasHeight}) {\(\ScaleLength{#3}\)};%
}
\newcommand{\WhiteScaleBar}[5]{%
    \draw[white,line width=1pt,line cap=butt]
    ({#1/\ScaleCanvasWidth},{1-#2/\ScaleCanvasHeight}) --
    ({(#1+(#3/#4*1000)*(#5))/\ScaleCanvasWidth},{1-#2/\ScaleCanvasHeight});%
    \ScaleBarLabel{#1+(#3/#4*1000)*(#5)/2}{#2}{#3}%
}

\newcommand{\ReferenceMosaicScaleLabels}{%
    {\renewcommand{\ScaleBarLabelShift}{2pt}\ScaleBarLabel{190}{232}{10}}
    \ScaleBarLabel{291}{232}{1}
}
\newcommand{\EggChamberScaleLabels}{%
    \ScaleBarLabel{31}{238}{\ActinOverviewBarUm}
    \ScaleBarLabel{31}{262}{\ActinOverviewBarUm}
    \ScaleBarLabel{513.5}{107}{\ActinDetailBarUm}
    \ScaleBarLabel{411.5}{356.5}{\ActinDetailBarUm}
    \ScaleBarLabel{513.5}{356.5}{\ActinDetailBarUm}
    \ScaleBarLabel{411.5}{461.5}{\ActinDetailBarUm}
    \ScaleBarLabel{513.5}{461.5}{\ActinDetailBarUm}
}

\newcommand{\DMExamplesScaleBars}{%
    \WhiteScaleBar{25}{334}{\DMImageBarUm}{\DMPixelSizeNm}{0.278269}
    \WhiteScaleBar{25}{407}{\DMImageBarUm}{\DMPixelSizeNm}{0.278269}
}
\newcommand{\DMComparisonScaleBars}{%
    {\renewcommand{\ScaleBarLabelShift}{3pt}%
        \WhiteScaleBar{5}{74}{\DMImageBarUm}{\DMPixelSizeNm}{0.152521}
    \WhiteScaleBar{51}{74}{\DMPSFBarUm}{\DMPixelSizeNm}{0.830831}}
    \WhiteScaleBar{6}{367}{\DMImageBarUm}{\DMPixelSizeNm}{0.272620}
    \WhiteScaleBar{6}{420}{\DMZoomBarUm}{\DMPixelSizeNm}{0.817859}
    \WhiteScaleBar{59}{420}{\DMZoomBarUm}{\DMPixelSizeNm}{0.817859}
    \WhiteScaleBar{6}{527}{\DMImageBarUm}{\DMPixelSizeNm}{0.272620}
    \WhiteScaleBar{6}{580}{\DMZoomBarUm}{\DMPixelSizeNm}{0.817859}
    \WhiteScaleBar{59}{580}{\DMZoomBarUm}{\DMPixelSizeNm}{0.817859}
}
\newcommand{\TiltScaleBars}{%
    \WhiteScaleBar{475}{1075}{\TiltImageBarUm}{\TiltPixelSizeNm}{594/800}
    \WhiteScaleBar{1418}{637}{\TiltZoomBarUm}{\TiltPixelSizeNm}{148.5/106}
    \WhiteScaleBar{1418}{1080}{\TiltPSFBarUm}{\TiltPixelSizeNm}{148.5/43}
}
\newcommand{\TiltPanelLabels}{%
    \PanelLabelLeft{0}{0}{a}
    \PanelLabel{975}{0}{b}
    \PanelLabelLeft{0}{505}{c}
}

\usepackage{float}
\usepackage{placeins} %
\usepackage[labelfont=bf, labelsep=period, font=small]{caption}
\usepackage{subcaption}

\usepackage{calc}
\usepackage{bbding}
\usepackage{booktabs}
\usepackage{xspace}
\usepackage{soul} %

\usepackage[dvipsnames]{xcolor}
\definecolor{green}{RGB}{0,100,0}        %
\definecolor{nmred}{RGB}{175,30,45}       %
\definecolor{nmgray}{RGB}{100,100,100}    %

\usepackage{xurl} %
\usepackage[
    colorlinks=true,
    linkcolor=nmred,
    citecolor=nmred,
    urlcolor=nmred
]{hyperref}
\usepackage{cleveref}
\usepackage[numbers,compress,sort]{natbib}
\usepackage{titlesec}

\titleformat{\section}
{\normalfont\bfseries\fontsize{11}{13}\selectfont\color{black}}
{}{0pt}{}

\titleformat{\subsection}
{\normalfont\bfseries\normalsize}
{}{0pt}{}

\titleformat{\subsubsection}[runin]
{\normalfont\itshape\normalsize}
{}{0pt}{}[.\enspace]

\titlespacing*{\section}{0pt}{8pt}{3pt}    %
\titlespacing*{\subsection}{0pt}{6pt}{2pt}
\titlespacing{\subsubsection}{0pt}{3pt}{0pt}

\usepackage{mdframed}
\newmdenv[
    linecolor=nmred,
    linewidth=1.2pt,
    topline=true, bottomline=false, leftline=false, rightline=false,
    innertopmargin=5pt, innerbottommargin=5pt,
    innerleftmargin=0pt, innerrightmargin=0pt,
    skipabove=6pt, skipbelow=6pt
]{abstractbox}

\usepackage{authblk}

\newcommand{\onlinemethods}{%
    \vspace{6pt}
    \noindent\rule{\columnwidth}{0.6pt}\vspace{2pt}
    {\fontsize{11}{13}\bfseries\selectfont ONLINE METHODS}\vspace{3pt}\par
}

\newcommand{\name}{DeepBlur\xspace}

\begin{document}
\setlength{\columnsep}{5mm}

\twocolumn[
    {
        \vspace*{2mm}

        {\fontsize{19}{22}\bfseries\selectfont\color{black}
            Blind spatially varying PSF estimation for fluorescence \\ microscopy\par
        }

        \vspace{5pt}

        \noindent
        {\normalfont\bfseries\fontsize{10}{12}\selectfont
            Minh Hai Nguyen\textsuperscript{1,2,\(\dag\)},
            Florian Sarron\textsuperscript{1,2,3,\(\dag\)},
            Paul Escande\textsuperscript{3},
            Sylvain Cantaloube\textsuperscript{2},
            Thomas Mangeat\textsuperscript{2} and
            Pierre Weiss\textsuperscript{1,2,3, \Envelope}
        }

        \vspace{3pt}

        {
            \fontsize{7.5}{9}\selectfont\color{nmgray}
            \textsuperscript{1} Université de Toulouse, CNRS, IRIT (Institut de Recherche en informatique de Toulouse), Toulouse, France \\
            \textsuperscript{2} Université de Toulouse, CNRS, CBI (Centre de Biologie Intégrative), Toulouse,  France\\
            \textsuperscript{3} Université de Toulouse, CNRS, IMT (Institut de Mathématiques de Toulouse), Toulouse, France \\
            \textsuperscript{\(\dag\)} The authors contributed equally.
            \Envelope\enspace \href{mailto:pierre.weiss@cnrs.fr}{pierre.weiss@cnrs.fr}
        }
        \par\vspace{5pt}
        \noindent\rule{\textwidth}{0.4pt}
        \vspace{2pt}

        \begin{abstractbox}
            {\fontsize{8.5}{11}\selectfont
                Optical aberrations degrade fluorescence microscopy images and measuring the underlying point spread function (PSF) often requires
                dedicated acquisitions or isolated emitters.
                Here we introduce \name{}, a neural network that estimates spatially varying two-dimensional PSFs directly from single images of
                biological structures.
                Trained entirely on physical simulations spanning aberrations, illumination and noise, it adapts to diverse imaging conditions
                without requiring user-supplied optical parameters.
                Using a deformable mirror, we acquire 1,290 images and assess blind PSF estimates against independent references.
                \name{} estimates agree closely with these references and improve deconvolution of experimentally aberrated images.
                Local PSF estimation also enables correction of spatially varying defocus and improves random-illumination super-resolution
                reconstruction in biological tissue.
            Available through DeepInv, a napari plugin and a web demonstration, \name{} can be readily integrated into computational microscopy workflows.}
        \end{abstractbox}

        \vspace{3pt}
        \noindent\rule{\textwidth}{0.4pt}
        \vspace{4pt}
}] %

Optical aberrations degrade contrast and resolution in fluorescence microscopy, impairing visual interpretation and downstream
quantification. Imperfections in the optical system and refractive-index variations within the specimen distort the wavefront,
producing blur that can vary across the field of view and change during an acquisition. The point spread function (PSF) is therefore often
specific to the sample, its position and the imaging conditions.

Hardware adaptive optics compensates these distortions by estimating the aberrated wavefront and applying a corrective phase, restoring
image quality at acquisition~\cite{ji2017adaptive}. Although effective, it requires specialized optical components and their integration into the
microscope.
Computational correction offers an accessible alternative to hardware-based approaches and can be applied retrospectively to acquired data, but
accurate knowledge of the PSF remains essential for deconvolution and many other reconstruction methods.

Several approaches exploit measurements for which the optical response is identifiable. Fluorescent beads and isolated blinking fluorophores
approximate point sources, allowing the PSF to be measured directly, even with field- and depth-dependent aberrations~\cite{liu2024universal}.
Other approaches exploit detector-array measurements in image scanning microscopy~\cite{koho2020two}, multiple images under shifted speckle
illumination~\cite{premillieu2024aberration}, or volumetric images combined with a microscope-specific forward model~\cite{matsuda2026phase}.
In contrast to these approaches, we seek to recover the local in-plane PSF from a single fluorescence image, without any dedicated measurements or
specialized acquisition protocol.

Deep learning has opened promising routes to this form of blind estimation. Networks infer local PSFs from extended microscopy images
for deconvolution~\cite{shajkofci2020spatially,debarnot2024deep} and aberration-aware super-resolution~\cite{qiao2024deep}, while image-restoration
networks trained on synthetic aberrations also improve experimentally acquired biological images~\cite{guo2025deep}. However, routine adoption of
blind PSF estimation appears limited. A likely obstacle is the lack of ready-to-use software that researchers can easily install and test on their
own images. A second obstacle is limited direct validation of PSF accuracy on real aberrations. This matters especially for simulation-trained
algorithms: as we show below, our earlier estimator~\cite{debarnot2024deep} loses accuracy on experimentally aberrated images despite its strong
performance on synthetic data.

Here we present \name{}, a blind estimator of smoothly varying two-dimensional (2D) PSFs from a single aberrated biological image, without requiring
isolated emitters or calibration targets. Our central contribution is an experimentally validated, ready-to-use PSF estimator: a neural network
released with pretrained weights and software that researchers can apply to their own images, including through an
\href{https://deepblur.mambo-utoulouse.fr}{online demo}. Its performance rests on a physical simulation and training procedure that spans
aberrations, noise, illumination and cutoff frequency across a broad range of conditions. The network is trained entirely on synthetic
data, using scalar diffraction theory and a Zernike parameterization of the wavefront, including all modes up to radial order 7. It enables automated
PSF estimation directly from the input image, without requiring explicit knowledge of the optical cutoff frequency or noise level.

We quantitatively assess PSF accuracy on 1{,}290 aberrated images acquired with a deformable mirror, alongside 42 flat-mirror images of
the same fields. Reference PSFs are recovered from paired non-aberrated and aberrated acquisitions, whereas \name{} receives only the aberrated
image. Despite this difference in available information, the estimated PSFs agree closely with the reference PSFs and enable effective restoration of
experimentally aberrated images.

The estimated in-plane PSF provides a reusable input to computational imaging. We demonstrate its utility for deconvolution with the Reconstruct
Anything Model (RAM)~\cite{terris2026reconstruct}, cell segmentation and random-illumination
super-resolution~\cite{Mangeat2021}. In this last application, speckle-based
statistical processing provides optical sectioning~\cite{Mangeat2021}, while our collection PSF estimates improve lateral reconstruction in
biological samples with sample-induced aberrations. The method is open-source, distributed as a \textit{napari} plugin and within the
\textit{DeepInverse} library~\cite{Tachella2025}.

\section{Results}

\subsection{\name{} estimates local PSFs from single images}

\begin{figure*}[htbp]
    \centering
    \includegraphics[width=\linewidth]{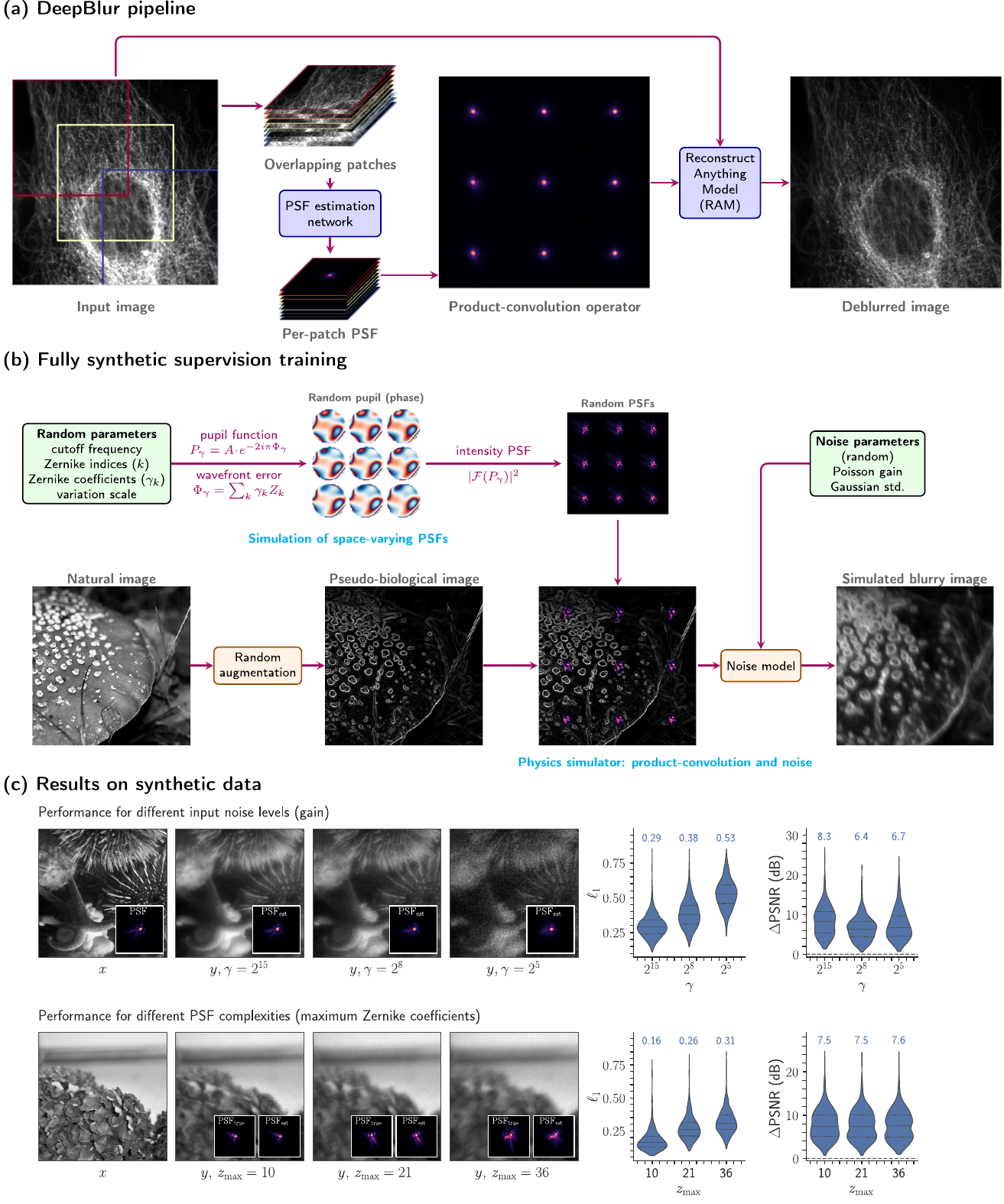}
    \caption{\textbf{Method overview.}
        (\textbf{a})~\name estimates a space-varying 2D PSF by processing overlapping image tiles and predicting the PSF at each tile center.
        The estimated PSF map can be used for downstream tasks such as deconvolution with the fine-tuned Reconstruct Anything Model (RAM).
        (\textbf{b})~\name is trained on simulated pairs of sharp and aberrated images with known PSFs, generated from a physics-based model of the
        microscope objective using scalar diffraction theory and Zernike aberration parameterization.
        (\textbf{c})~Synthetic PSF estimates (insets) and deconvolution performance across Poisson noise levels (\(\gamma\), top) and maximum Zernike
        indices (\(z_\mathrm{max}\), bottom).
        Violin plots show the PSNR gain after deconvolution (\(\Delta\mathrm{PSNR}\)) and the \(\ell^1\) distance between estimated and ground-truth
    PSFs. PSNR gains are relative to the blurred input; numbers above the violins are medians over 1024 images.}
    \label{fig:method}
\end{figure*}

The central output of \name is not a restored image but the optics response: a smoothly space-varying 2D PSF estimated directly from a
single aberrated image (Fig.~\ref{fig:method}a).
A network reads the input through overlapping tiles and predicts, at each tile centre, the local PSF; assembling these local estimates yields a
space-varying blur operator, represented as a product-convolution expansion~\cite{Escande2017}.
The network combines a residual convolutional encoder, a multiscale context module and a head that predicts PSF pixel values directly
(Extended Data Fig.~\ref{edfig:architecture}).
The optical parameters such as the numerical aperture or pixel size need not be specified.
A separate network estimates the noise parameters used for reconstruction; no calibration acquisition is required.

An estimated PSF can be supplied to computational imaging methods that use an explicit blur model, including deconvolution and structured- or
speckle-illumination super-resolution. We exploit this throughout: for deblurring we pair the estimate with a fine-tuned Reconstruct
Anything Model (RAM)~\cite{terris2026reconstruct}, while for super-resolution we supply the estimated collection PSF to a random-illumination microscopy (RIM)
algorithm~\cite{Mangeat2021}.
The PSF estimator is thus a reusable module that provides an interpretable representation of the aberration. It can be inspected independently of the
restored image.

The estimator is trained entirely on synthetic data (Fig.~\ref{fig:method}b).
The clean training images come from a large pool of natural images drawn from diverse datasets, with random augmentations chosen to mimic
fluorescence-like structures. We then simulate aberrated images with a physics-based forward model of an incoherent microscope: a Zernike wavefront defines
a pupil function, and the squared magnitude of its Fourier transform gives the intensity PSF; this PSF is
applied to the illuminated image, which is then corrupted by a realistic Poisson--Gaussian noise model.
Aberrations, noise, illumination and cutoff frequency are drawn at random for every training example, exposing the network to a broad range of
imaging conditions.
Our aim is to cover a family of plausible optical conditions and image structures during training, so that variation within these simulated ranges is
learned directly rather than left to extrapolation from a narrow setting~\cite{adaptive_networks_gossard_2022}.
Examples of simulated PSFs are shown in Extended Data Fig.~\ref{edfig:example_PSFs}.
Full details are given in Online Methods.

We first evaluated the method on synthetic data, where the ground-truth PSF is known exactly.
We assessed both PSF-estimation error and the gain in image quality after deconvolution with RAM.
Restoration yielded substantial improvements in image fidelity across the tested noise levels and aberration orders, with median PSNR gains of
6.4--8.3\,dB over the blurred input (Fig.~\ref{fig:method}c).
Including higher-order Zernike modes during training substantially improved PSF estimation for higher-order aberrations without compromising accuracy
for lower-order aberrations, supporting the use of a single model (Extended Data Fig.~\ref{edfig:zernike_generalization}).

\subsection{Experimental validation with a deformable mirror}

\begin{figure*}[htbp]
    \centering
    \FigureWithScaleBars{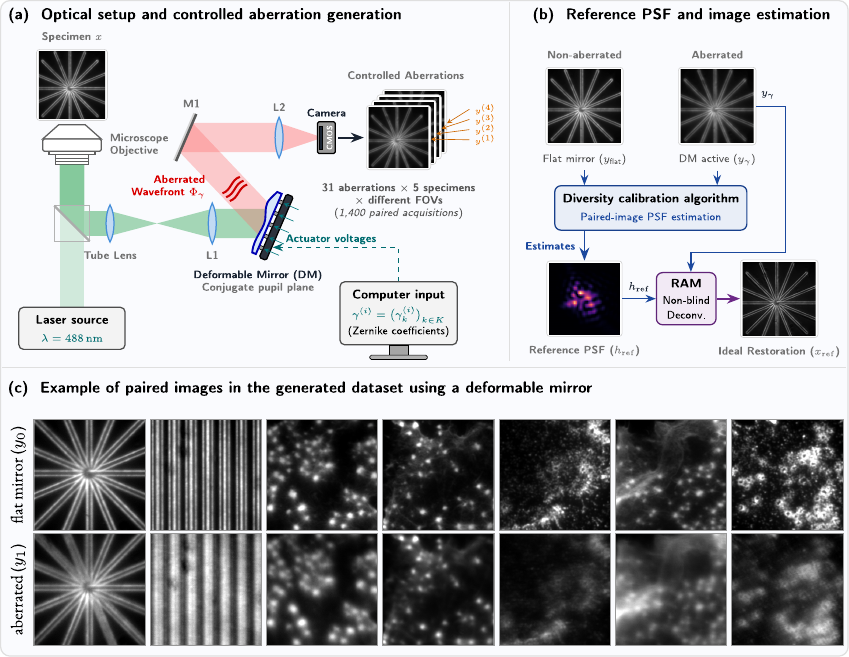}{543.638}{420.651}{\DMExamplesScaleBars}
    \caption{\textbf{\name experimental validation with a deformable mirror.}
        (\textbf{a})~We recorded each field of view with a flat mirror and under multiple mirror-induced aberrations, yielding 1\,290 (flat-mirror,
        aberrated) image pairs for experimental validation.
        (\textbf{b})~From each pair, a reference PSF is estimated via a
        dedicated algorithm, and used both to evaluate \name and to produce
        an image deconvolved in ideal conditions.
    (\textbf{c})~Examples of paired non-aberrated (top) and aberrated (bottom) acquisitions.}
    \label{fig:dm}
\end{figure*}

To test whether estimates trained on simulated data transfer to real optical aberrations, we built an experimental benchmark using a deformable
mirror (Fig.~\ref{fig:dm}a).
The mirror was positioned in a plane conjugate to the objective pupil. At each field of view, we changed its shape to image the same sample region
under different optical aberrations.
We recorded 1{,}290 aberrated images using 42 different fields of view under 31 distinct mirror settings.
In addition, 42 flat-mirror images of the same fields were acquired to provide a reference.
Each aberrated image was paired with the flat-mirror image of its field for reference-PSF estimation.
The samples differed in emission wavelength and structure.
The database is available at \url{https://zenodo.org/records/24304044}.

Evaluating a blind estimate requires a reference PSF for each aberrated image. We
obtained one for each of the 1{,}290 aberrated images by an independent procedure: given the flat-mirror image obtained with a near ideal Airy
pattern PSF, and the aberrated image
of the same object, we recovered the aberrated PSF using a calibration diversity algorithm developed in~\cite{qdo2026phase} (Fig.~\ref{fig:dm}b).
This reference exploits information that \name never accesses---a second, non-aberrated acquisition---so it serves as a stringent, informed benchmark
for the single-image blind estimate,
obtained through an entirely different route.
Reference PSFs for representative mirror states are shown in Extended Data Fig.~\ref{edfig:estimated_PSFs_DM}.

\begin{figure*}[!tp]
    \centering
    \FigureWithScaleBars{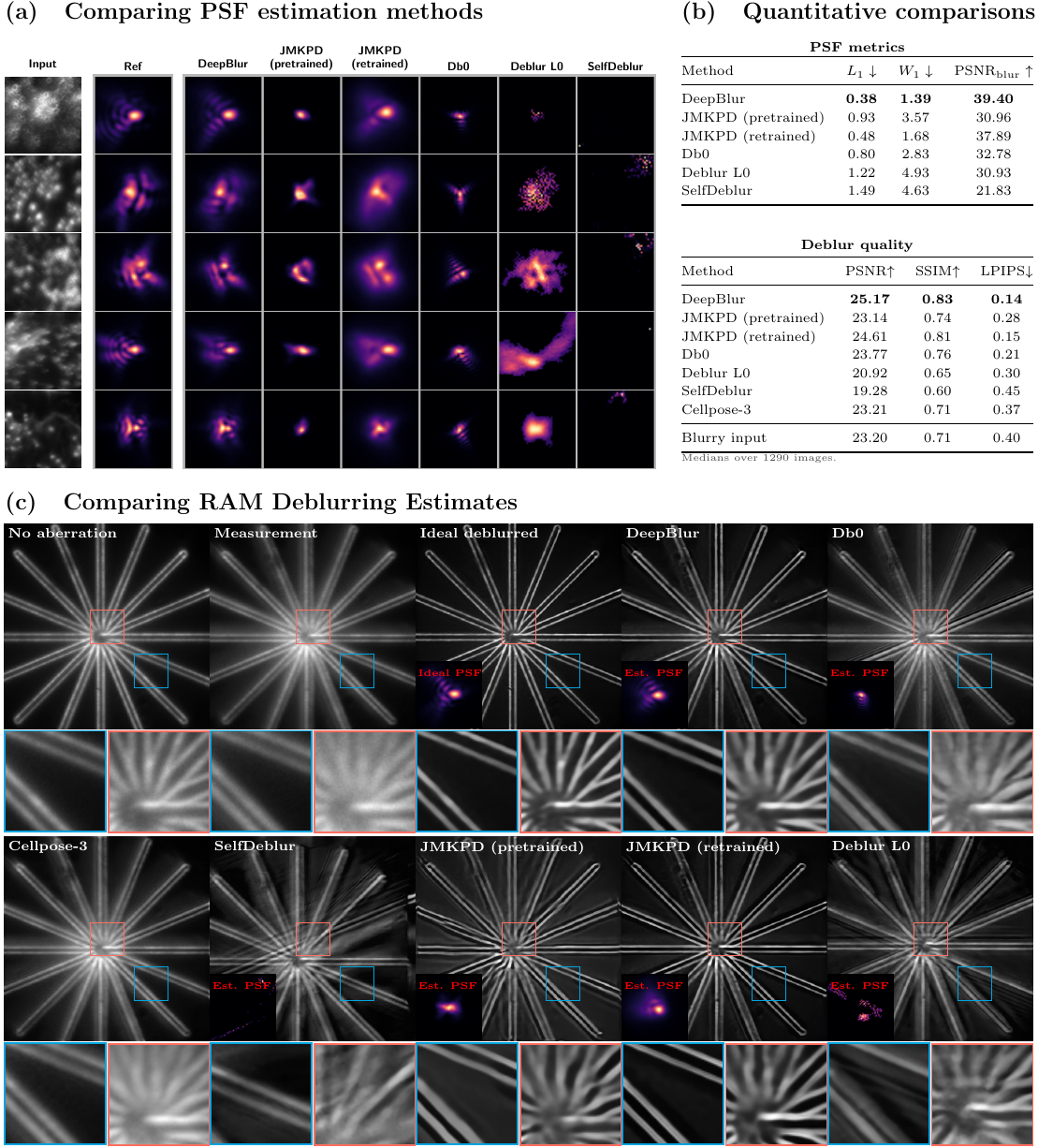}{531.03}{586.949}{\DMComparisonScaleBars}
    \caption{\textbf{Qualitative and quantitative comparisons on the deformable mirror dataset.}
        (\textbf{a})~Qualitative comparison with competing methods: the aberrated
        image (top), the reference PSF and PSFs produced by each
        method~\cite{carbajal2023blind,debarnot2024deep,ipol.2019.243,pan2014deblurring,ren2020neural}
        (bottom). JMKPD~\cite{carbajal2023blind} is shown with both its original pretrained weights and weights retrained using our procedure;
        retraining improves agreement with the reference PSF.
        Db0~\cite{debarnot2024deep}, L0~\cite{pan2014deblurring} and
        SelfDeblur~\cite{ren2020neural} produce unreliable estimates.
        (\textbf{b})~Across PSF-estimation and deblurring metrics, \name{} outperforms competing blind methods; the same fine-tuned RAM model is
        used for all methods, so only the estimated PSF differs.
    (\textbf{c})~Qualitative comparisons between the fine-tuned RAM deblurring method with PSF estimates produced by competing methods.}
    \label{fig:dm2}
\end{figure*}

Given only a single experimentally acquired aberrated image, \name{} recovered PSFs that closely matched the independent references
(Fig.~\ref{fig:dm2}a). Across the 1{,}290 aberrated images, it achieved the best median values on all PSF measures (Fig.~\ref{fig:dm2}b). The
importance of our training procedure is made clear using the JMKPD architecture~\cite{carbajal2023blind}.
With its original pretrained weights, JMKPD performed poorly; retraining the same architecture with our simulation and training procedure nearly
halved its median error, making it the strongest competitor to \name{}.
This improvement with the architecture held fixed highlights the importance of training for transfer to real aberrations, while \name{}'s further
gain shows that architecture also matters. Our earlier synthetic-only estimator Db0~\cite{debarnot2024deep} was used as published;
L0~\cite{pan2014deblurring} and SelfDeblur~\cite{ren2020neural} were optimized separately for each image without pretraining. All three yielded
substantially less accurate PSFs.

We next evaluated how well a fine-tuned RAM model~\cite{terris2026reconstruct} restored the experimentally aberrated images using the estimated PSF
from each method. Keeping the reconstruction model fixed also provided an image-based assessment of the PSF estimates. \name's
estimates achieved the best restoration on every image-quality metric (Fig.~\ref{fig:dm2}b), surpassing the aberrated input and the
segmentation-oriented restoration model
Cellpose3~\cite{stringer2025cellpose3}, and approaching the quality obtained with the
reference PSF itself. Qualitatively, on the Argolight star target, fine radial structures lost to aberration
were restored when RAM was driven by \name's PSF, yet remained blurred or
corrupted with the alternatives (Fig.~\ref{fig:dm2}c).
Across the 1{,}290 aberrated images, RAM deconvolution with \name-estimated PSFs improved PSNR, SSIM, and LPIPS in over 80\%, 90\%, and 98\% of
cases, respectively, compared with the aberrated inputs (Extended Data Fig.~\ref{edfig:dm_violins}).

These results show that a network trained entirely on simulations can infer, from a single aberrated image, a reliable PSF useful for correcting real
experimental data.

\subsection{\name{} estimates space-varying defocus}

\begin{figure*}[!tp]
    \centering
    \FigureWithScaleBars[.95\linewidth]{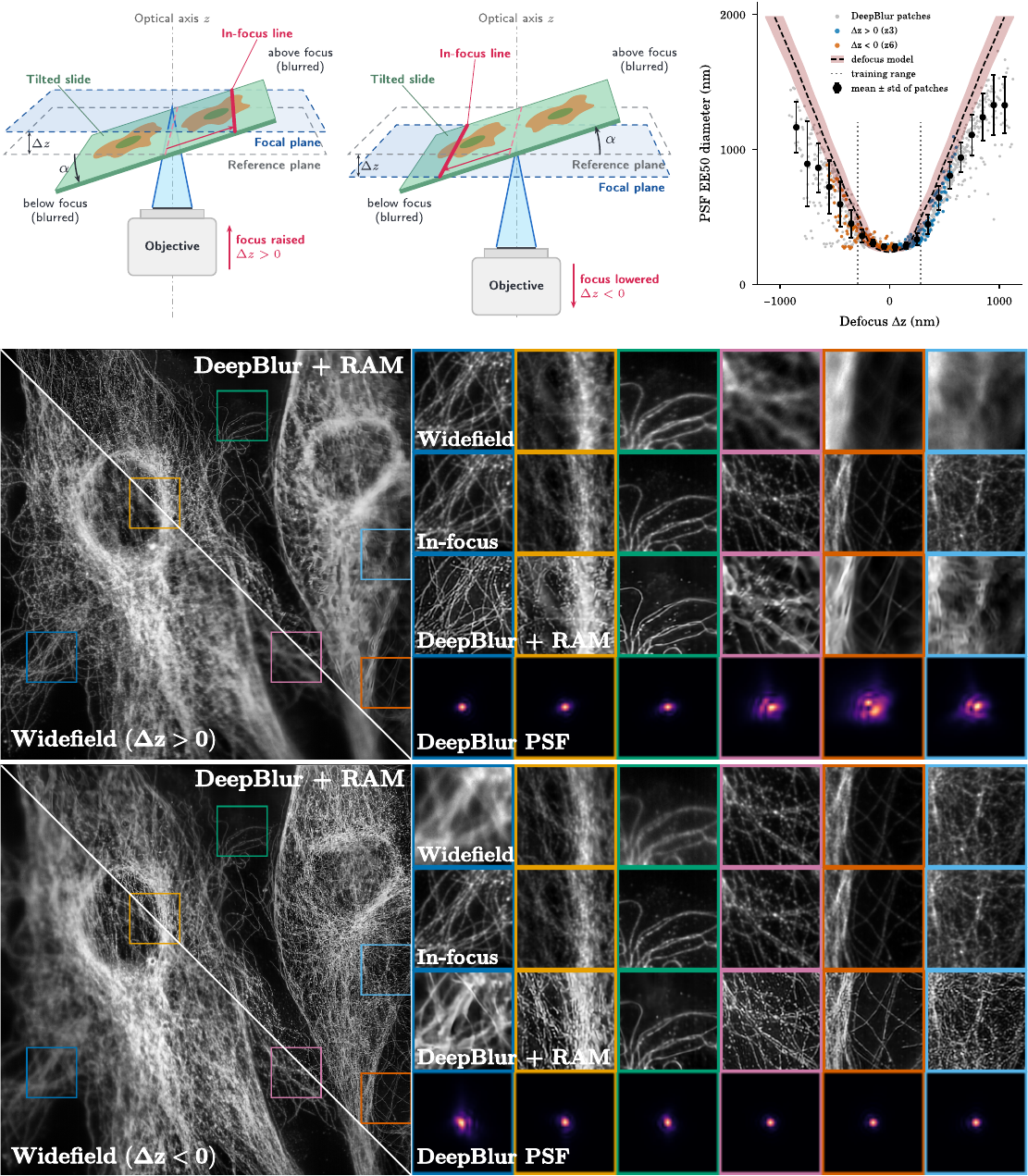}{1486}{1700}{\TiltScaleBars\TiltPanelLabels}
    \caption{\textbf{Spatially varying PSF estimation and restoration in a tilted biological specimen.}
        Texas Red-X phalloidin-labelled BPAECs on a tilted prepared slide were imaged with an NA \(1.49\) objective.
        (\textbf{a})~Raising (left) or lowering (right) the axial focal position shifts the in-focus region across the tilted specimen.
        (\textbf{b})~Width of the \name{} PSFs (EE50: diameter enclosing 50\% of the PSF energy) versus the defocus \(\Delta z\) of each patch,
        measured independently from a nine-plane \(z\)-stack (Methods). Grey dots, all 1{,}089 patches; coloured dots, the patches shown in
        (\textbf{c}); black, average width (\(\pm\) s.d.). Dashed line and shaded band, theoretical width of an ideal PSF at each defocus. Dotted
        lines, defocus range of the training set (\(\pm 287\)\,nm).
        (\textbf{c})~Positive (upper block) and negative (lower block) focal offsets \(\Delta z\) relative to the reference plane.
        In each block, the overview combines the widefield acquisition (below the diagonal) and the reconstruction obtained with \name PSFs and RAM
        (above the diagonal). Coloured boxes identify the enlarged regions shown to the right.
        Rows show the widefield image, the local in-focus reference from a \(z\)-stack, the \name{} + RAM reconstruction and the estimated local PSF.
    In-focus views provide structural references for the reconstructions.}
    \label{fig:space_varying}
\end{figure*}

\name{} estimates PSFs that vary across the field of view from a single image.
We tested this by deliberately tilting a prepared slide relative to the focal plane.
The slide contained a layer of bovine pulmonary artery endothelial cells (BPAECs) labelled with Texas Red-X phalloidin, which we imaged with an
NA~1.49 objective (Fig.~\ref{fig:space_varying}a).
Across the \(78\,\mu\mathrm{m}\) field of view, defocus therefore varies linearly with lateral position, resulting in progressively broader PSFs on
either side of the in-focus line. Changing the axial focus position shifts the in-focus line across the field of view.

To estimate the true defocus, we located the best-focus plane of each region by keeping the sharpest patch in a nine-plane \(z\)-stack (Methods).
We then applied \name{} patch by patch to each plane separately, yielding one PSF per region and plane.
Across the 1{,}089 resulting PSFs, each estimated from a single image, PSF width (diameter of the smallest circle enclosing 50\% of the PSF
intensity) was strongly correlated with the absolute axial distance to best focus: PSFs were narrowest at best focus, close to the diffraction limit
(\(277 \pm 19\)\,nm, versus 238\,nm for an ideal in-focus PSF), and broadened on both sides.
Within the defocus range covered at training time (\(\pm 287\)\,nm), the widths were consistent with the reference defocus with less than 12\% error
for half of the patches; beyond it, the defocus width estimated by \name{} kept increasing but stayed below reality with a median error of 22\%.
Larger defocus values were deliberately excluded from training because the attenuation of high spatial frequencies makes PSF estimation increasingly
difficult. Nevertheless, the estimated PSF widths continued to increase with defocus beyond the training range, suggesting that the network
generalizes at least qualitatively to larger defocus values.

Fig.~\ref{fig:space_varying}c illustrates this behaviour at two focal positions, above (\(\Delta z > 0\)) and below
(\(\Delta z < 0\)) the reference plane. Each overview combines the widefield image (below the diagonal) with its restoration by RAM
using the \name{} PSF map (above the diagonal). For six regions spread across the field, the enlarged views show the widefield image, an
in-focus reference taken from the \(z\)-stack, the restoration and the local PSF.
PSFs are compact near the in-focus line and broaden away from it.
The pattern reverses between the two focal positions.
In defocused regions, the restoration recovers filaments that match those of the in-focus references, showing that the local PSF estimates enable
lateral restoration across a specimen whose distance to the focal plane varies within the field of view.

\subsection{PSF estimation improves super-resolved reconstruction in random illumination microscopy}

\begin{figure*}[!tp]
    \centering
    \includegraphics[width=\textwidth,height=.85\textheight,keepaspectratio]{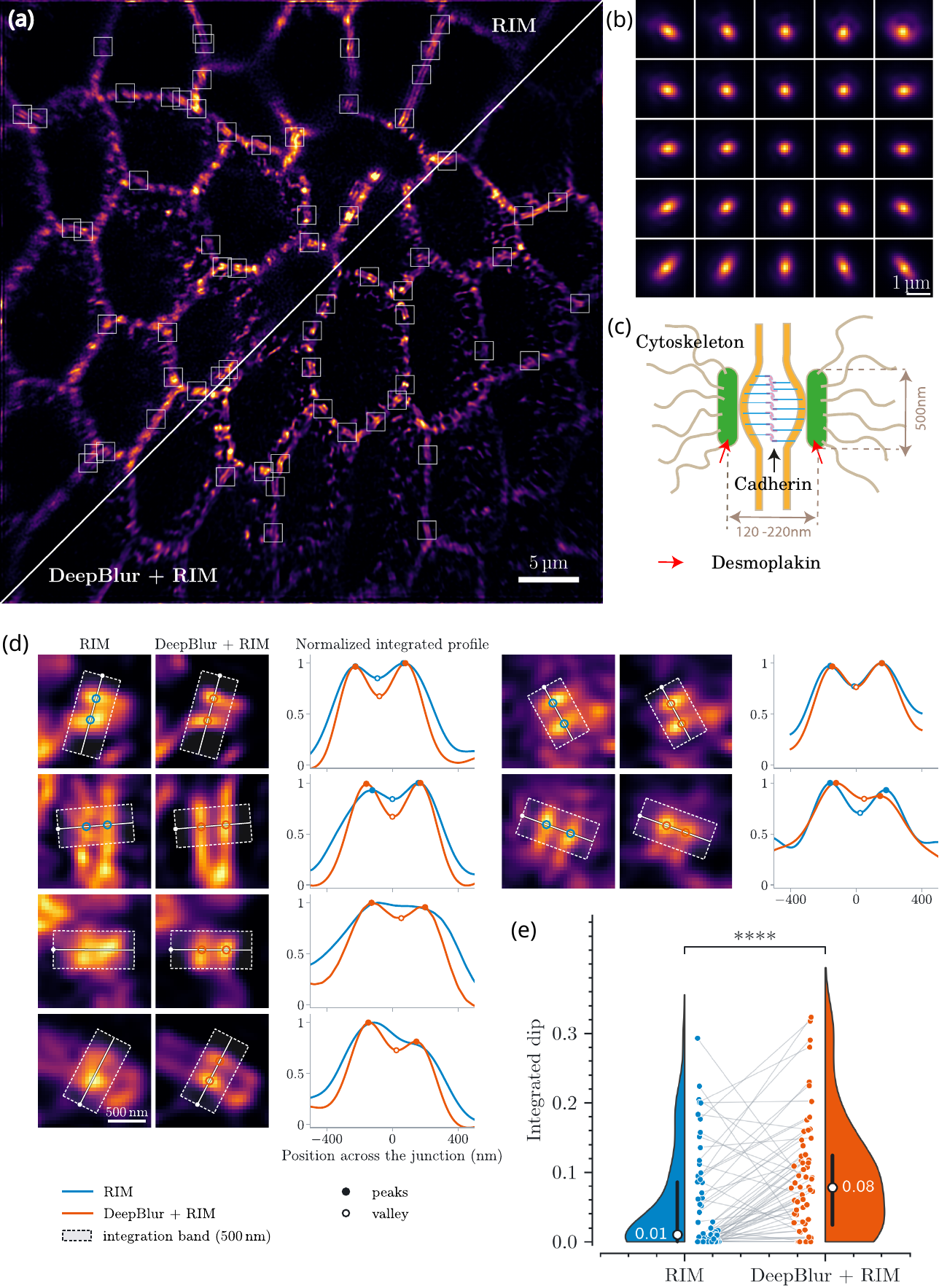}
    \caption{{\textbf{Improved separation of desmoplakin structures in RIM reconstructions using \name-estimated PSFs.}
            Desmoplakin-immunolabelled mouse intestinal epithelium imaged using 2D-RIM with a pupil-plane phase mask.
            (\textbf{a})~Overview comparing reconstructions of the same acquisitions using the theoretical collection PSF (RIM, above the diagonal)
            and local PSFs estimated by \name{} (\name{} + RIM, below the diagonal). Boxes mark analysed regions.
            (\textbf{b})~Spatial grid of estimated collection PSFs.
            (\textbf{c})~Schematic of desmoplakin localisation on opposing sides of a desmosomal junction.
            (\textbf{d})~Corresponding enlarged regions and normalized intensity profiles across the junctions, comparing RIM (blue)
            and \name{} + RIM (orange). Dashed rectangles indicate the 500\,nm-wide integration bands; filled and open circles mark
            peaks and the intervening minimum, respectively.
            (\textbf{e})~Distribution of the interpeak dip measured from the integrated profiles. Lines connect measurements from the same object
    in the two reconstructions.}}
    \label{fig:RIM}
\end{figure*}

We next tested whether PSFs estimated by \name{} could also improve super-resolution reconstruction by supplying them to AlgoRIM~\cite{Mangeat2021},
the reconstruction algorithm used in random illumination microscopy (RIM; Fig.~\ref{fig:RIM}).
RIM reconstructs a super-resolved image from a sequence of speckle-illuminated acquisitions using the illumination statistics and the collection
PSF. Its optical sectioning and robustness to excitation aberrations make it well suited to imaging within biological tissue,
but reconstruction quality still depends on an appropriate collection PSF. RIM therefore allows us to test whether local 2D PSF estimates improve
lateral reconstruction within thick specimens.

For each spatial region, we applied \name{} to the averaged speckle acquisitions to estimate the local collection PSF.
We then supplied the local collection PSF to AlgoRIM and compared reconstructions of the same acquisitions using either the theoretical collection
PSF or the local PSFs estimated by \name.

{We examined} desmoplakin-immunolabelled mouse intestinal epithelium using 2D-RIM with a pupil-plane phase mask
(Fig.~\ref{fig:RIM}). The specimen was also used in the study of Mazzella et al.~\cite{mazzella2024extended}.
{Desmoplakin forms paired plaques on opposing sides of desmosomal junctions. Reported mean plaque-to-plaque distances between its C-terminal domains
    range from approximately 120 to 220\,nm depending on the isoform~\cite{ainslie2025desmoplakin}, making these structures a relevant biological
    benchmark for super-resolution imaging.
    Using \name-estimated PSFs improved the separation of the paired structures: among 64 analysed regions, the number of intensity profiles
    exhibiting an interpeak minimum increased from 43 (67\%) with the theoretical PSF to 54 (84\%) with \name-estimated PSFs.
    The median dip increased from 0.01 (interquartile range, 0--0.09) to 0.08 (0.03--0.15), demonstrating improved discrimination of the two plaques in
the super-resolved reconstruction (Fig.~\ref{fig:RIM}d,e).}

In a complementary experiment, \name{} also improved resolution of actin cables imaged with RIM deep within a \textit{Drosophila} egg chamber
(Extended Data Fig.~\ref{edfig:rim_egg_chamber}).

\subsection{\name{} for other downstream applications}
On synthetically degraded Cellpose images, restoration using \name-estimated PSFs also improved segmentation across five tested models (Extended
Data Fig.~\ref{edfig:segmentation}), demonstrating how PSF estimation can help image analysis tasks.

We tested the pipeline on the public dataset BioSR~\cite{qiao2021biosr}, in which each widefield image is paired with a super-resolved image of the
same cells. Restoring the widefield images with \name{} + RAM brought them closer to these higher-resolution references for all five tested image
groups. As expected for images acquired under identical optical conditions, the PSF estimates remained stable across the nine noise levels of the
dataset, with FWHMs ranging from 197 to 209 nm for a given field (Extended Data Fig.~\ref{edfig:biosr_gains}).

We also applied \name{} and RAM independently to the lateral planes of a spinning-disk confocal zebrafish stack (Extended Data
Fig.~\ref{edfig:zebrafish}). The restored images show improved lateral definition of membrane structures within the volumetric acquisition.

\section*{Discussion}

\name{} estimates local two-dimensional PSFs directly from fluorescence images of biological specimens, without relying on isolated point emitters or
dedicated calibration acquisitions.
Its performance rests on a physics-based training procedure and is validated quantitatively on experimental data: although trained entirely on
synthetic data, the estimator recovers PSFs from experimentally acquired images that agree closely with independently obtained references.
These comparisons assess the estimated PSF itself, complementing the improvements in image reconstruction
demonstrated across the experimental applications.
PSF estimation is fully automated, requiring neither user-supplied optical parameters nor prior knowledge of the noise level.

Image-based aberration estimation builds on earlier methods for local PSF prediction and spatially varying
deconvolution~\cite{shajkofci2018semi,shajkofci2020spatially,qiao2024deep}, alongside approaches that learn to restore aberrated images
directly~\cite{guo2025deep}.
Guo et al. demonstrate restoration of 3D volumes and report improved performance when the simulated aberrations match those of the sample. Their
workflow trains a separate model for each microscope and sample type using relatively well-resolved shallow volumes.
We instead train one PSF estimator across a broad simulated range and apply the same weights throughout our experimental datasets, allowing use when
sample-specific training data are unavailable.
A complementary route recovers detailed PSF models from fluorescent beads or isolated blinking molecules~\cite{liu2024universal}.
In particular, in-situ single-molecule methods can capture sample-induced aberrations at the location of the emitters.
\name{} addresses a different acquisition setting: it exploits the spatial information contained in fluorescence images of biological specimens,
without requiring isolated emitters or their temporal separation. This enables PSF estimation from previously acquired images and from experiments
whose labelling and acquisition protocols were not designed for optical calibration.

The comparison with JMKPD~\cite{carbajal2023blind} highlights the importance of the training procedure: retraining the same architecture transformed
inaccurate estimates into PSFs close to the experimental references.
Architecture also matters, as several other retrained models remained less accurate and our architecture provided a further improvement. These
observations support the joint design of the neural network architecture and its training distribution.

The two-dimensional PSFs estimated here are useful for correcting lateral blur in near-planar biological structures and optically sectioned images.
The tilted-cell experiment demonstrates correction of spatially varying defocus, while coupling \name{} to AlgoRIM shows that an image-derived PSF
can improve a separate reconstruction algorithm, including within biological tissue.
These applications illustrate the value of an explicit PSF that can be inspected and reused across computational imaging workflows.
They do not require full volumetric reconstruction, although a single 2D kernel cannot generally describe the mixing of fluorescence from multiple
depths in a thick specimen.

Beyond restoration, \name{} could provide a simple tool for routine microscope monitoring.
Established metrology protocols use PSF measurements to assess resolution and identify changes in optical
performance~\cite{cole2011measuring,faklaris2022quality}.
Estimating PSFs from a few images of extended specimens could offer a rapid, low-cost check for broadening, asymmetry or variation across the field,
without preparing dedicated point-source samples.
Repeated measurements of stable specimens under comparable acquisition conditions could help monitor changes after optical adjustments or over time.
This would complement quantitative calibration protocols with an accessible screening step, particularly in laboratories with limited optical expertise.

The same principle could support closed-loop adaptive optics, using image-derived PSFs to guide adjustments of a corrective element~\cite{ji2017adaptive}.
A further direction is to infer a pupil representation, such as Zernike coefficients, from the estimated PSF.
With suitable treatment of ambiguities and adapted training, such a representation could support both wavefront correction and prediction of a
three-dimensional PSF by propagation in a homogeneous medium.
These extensions would require dedicated validation.
For the applications demonstrated here, the open-source implementation in \href{https://deepinv.org/index.html}{DeepInverse}~\cite{Tachella2025}, the
\href{https://napari-hub.org/plugins/deepblur-napari.html}{napari plugin} and the \href{https://deepblur.mambo-utoulouse.fr}{web demonstration} make
image-based PSF estimation readily accessible within existing microscopy workflows.

\FloatBarrier
\bibliography{references}   %

@article{debarnot2024deep,
  title     = {Deep-blur: Blind identification and deblurring with convolutional neural networks},
  author    = {Debarnot, Valentin and Weiss, Pierre},
  journal   = {Biological Imaging},
  volume    = {4},
  pages     = {e13},
  year      = {2024},
  publisher = {Cambridge University Press}
}

@article{adaptive_networks_gossard_2022,
  title     = {Training adaptive reconstruction networks for blind inverse problems},
  author    = {Gossard, Alban and Weiss, Pierre},
  journal   = {SIAM Journal on Imaging Sciences},
  volume    = {17},
  number    = {2},
  pages     = {1314--1346},
  year      = {2024},
  publisher = {SIAM}
}

@article{Tachella2025,
  title     = {{DeepInverse}: A {Python} package for solving imaging inverse problems with deep learning},
  volume    = {10},
  issn      = {2475-9066},
  doi       = {10.21105/joss.08923},
  number    = {115},
  journal   = {Journal of Open Source Software},
  publisher = {The Open Journal},
  author    = {Tachella, Juli{\'a}n and Terris, Matthieu and Hurault, Samuel and Wang, Andrew and Davy, Leo and Scanvic, J{\'e}r{\'e}my and Sechaud, Victor and Vo, Romain and Moreau, Thomas and Davies, Thomas and Chen, Dongdong and Laurent, Nils and Monroy, Brayan and Dong, Jonathan and Hu, Zhiyuan and Nguyen, Minh-Hai and Sarron, Florian and Weiss, Pierre and Escande, Paul and Massias, Mathurin and Modrzyk, Thibaut and Levac, Brett and Liaudat, Tob{\'i}as I. and Song, Maxime and Hertrich, Johannes and Neumayer, Sebastian and Schramm, Georg},
  year      = {2025},
  month     = nov,
  pages     = {8923}
}

@article{shajkofci2020spatially,
  title     = {Spatially-variant {CNN}-based point spread function estimation for blind deconvolution and depth estimation in optical microscopy},
  author    = {Shajkofci, Adrian and Liebling, Michael},
  journal   = {IEEE Transactions on Image Processing},
  volume    = {29},
  pages     = {5848--5861},
  year      = {2020},
  publisher = {IEEE}
}

@article{Escande2017,
  author  = {Escande, Paul and Weiss, Pierre},
  title   = {Approximation of integral operators using product-convolution expansions},
  journal = {Journal of Mathematical Imaging and Vision},
  year    = {2017},
  day     = {01},
  volume  = {58},
  number  = {3},
  pages   = {333--348},
  issn    = {1573-7683},
  doi     = {10.1007/s10851-017-0714-8}
}

@article{liu2024universal,
  title     = {Universal inverse modeling of point spread functions for {SMLM} localization and microscope characterization},
  author    = {Liu, Sheng and Chen, Jianwei and Hellgoth, Jonas and M{\"u}ller, Lucas-Raphael and Ferdman, Boris and Karras, Christian and Xiao, Dafei and Lidke, Keith A and Heintzmann, Rainer and Shechtman, Yoav and others},
  journal   = {Nature Methods},
  volume    = {21},
  number    = {6},
  pages     = {1082--1093},
  year      = {2024},
  publisher = {Nature Publishing Group US New York}
}

@article{noll1976zernike,
  title     = {{Zernike} polynomials and atmospheric turbulence},
  author    = {Noll, Robert J},
  journal   = {Journal of the Optical Society of America},
  volume    = {66},
  number    = {3},
  pages     = {207--211},
  year      = {1976},
  publisher = {Optical Society of America}
}

@article{Lakshminarayanan10042011,
  author    = {Vasudevan Lakshminarayanan and Andre Fleck},
  title     = {{Zernike} polynomials: a guide},
  journal   = {Journal of Modern Optics},
  volume    = {58},
  number    = {7},
  pages     = {545--561},
  year      = {2011},
  publisher = {Taylor \& Francis},
  doi       = {10.1080/09500340.2011.554896}
}

@article{Jin:18,
  author    = {Yuncheng Jin and Yiye Zhang and Lejia Hu and Haiyang Huang and Qiaoqi Xu and Xinpei Zhu and Limeng Huang and Yao Zheng and Hui-Liang Shen and Wei Gong and Ke Si},
  journal   = {Optics Express},
  number    = {23},
  pages     = {30162--30171},
  publisher = {Optica Publishing Group},
  title     = {Machine learning guided rapid focusing with sensor-less aberration corrections},
  volume    = {26},
  month     = {Nov},
  year      = {2018},
  doi       = {10.1364/OE.26.030162}
}

@inproceedings{shajkofci2018semi,
  title        = {Semi-blind spatially-variant deconvolution in optical microscopy with local point spread function estimation by use of convolutional neural networks},
  author       = {Shajkofci, Adrian and Liebling, Michael},
  booktitle    = {Proceedings of the IEEE International Conference on Image Processing (ICIP)},
  pages        = {3818--3822},
  year         = {2018},
  organization = {IEEE}
}

@inproceedings{he2016deep,
  title     = {Deep residual learning for image recognition},
  author    = {He, Kaiming and Zhang, Xiangyu and Ren, Shaoqing and Sun, Jian},
  booktitle = {Proceedings of the IEEE Conference on Computer Vision and Pattern Recognition (CVPR)},
  pages     = {770--778},
  year      = {2016}
}

@article{ulyanov2016instance,
  title   = {Instance normalization: The missing ingredient for fast stylization},
  author  = {Ulyanov, Dmitry and Vedaldi, Andrea and Lempitsky, Victor},
  journal = {arXiv preprint arXiv:1607.08022},
  year    = {2016}
}

@article{hendrycks2016gaussian,
  title   = {{Gaussian} error linear units ({GELUs})},
  author  = {Hendrycks, Dan and Gimpel, Kevin},
  journal = {arXiv preprint arXiv:1606.08415},
  year    = {2016}
}

@article{chen2017rethinking,
  title   = {Rethinking atrous convolution for semantic image segmentation},
  author  = {Chen, Liang-Chieh and Papandreou, George and Kokkinos, Iasonas and Murphy, Kevin and Yuille, Alan L},
  journal = {arXiv preprint arXiv:1706.05587},
  year    = {2017}
}

@inproceedings{wu2018group,
  title     = {Group normalization},
  author    = {Wu, Yuxin and He, Kaiming},
  booktitle = {Proceedings of the European Conference on Computer Vision (ECCV)},
  pages     = {3--19},
  year      = {2018}
}

@inproceedings{deng2009imagenet,
  title        = {{ImageNet}: A large-scale hierarchical image database},
  author       = {Deng, Jia and Dong, Wei and Socher, Richard and Li, Li-Jia and Li, Kai and Fei-Fei, Li},
  booktitle    = {Proceedings of the IEEE Conference on Computer Vision and Pattern Recognition (CVPR)},
  pages        = {248--255},
  year         = {2009},
  organization = {IEEE}
}

@inproceedings{kingma2017adammethodstochasticoptimization,
  title     = {Adam: A method for stochastic optimization},
  author    = {Kingma, Diederik P. and Ba, Jimmy},
  booktitle = {International Conference on Learning Representations (ICLR)},
  year      = {2015}
}

@inproceedings{terris2026reconstruct,
  title     = {{Reconstruct Anything Model}: a lightweight general model for computational imaging},
  author    = {Matthieu Terris and Samuel Hurault and Maxime Song and Juli{\'a}n Tachella},
  booktitle = {International Conference on Learning Representations (ICLR)},
  year      = {2026}
}

@article{Mangeat2021,
  author    = {Mangeat, Thomas
               and Labouesse, Simon
               and Allain, Marc
               and Negash, Awoke
               and Martin, Emmanuel
               and Gu{\'e}nol{\'e}, Aude
               and Poincloux, Renaud
               and Estibal, Claire
               and Bouissou, Ana{\"i}s
               and Cantaloube, Sylvain
               and Vega, Elodie
               and Li, Tong
               and Rouvi{\`e}re, Christian
               and Allart, Sophie
               and Keller, Debora
               and Debarnot, Valentin
               and Wang, Xia Bo
               and Michaux, Gr{\'e}goire
               and Pinot, Mathieu
               and Le Borgne, Roland
               and Tournier, Sylvie
               and Suzanne, Magali
               and Idier, J{\'e}r{\^o}me
               and Sentenac, Anne},
  title     = {Super-resolved live-cell imaging using random illumination microscopy},
  journal   = {Cell Reports Methods},
  year      = {2021},
  month     = {May},
  day       = {24},
  publisher = {Elsevier},
  volume    = {1},
  number    = {1},
  pages     = {100009},
  issn      = {2667-2375},
  doi       = {10.1016/j.crmeth.2021.100009}
}

@article{cole2011measuring,
  title     = {Measuring and interpreting point spread functions to determine confocal microscope resolution and ensure quality control},
  author    = {Cole, Richard W and Jinadasa, Tushare and Brown, Claire M},
  journal   = {Nature Protocols},
  volume    = {6},
  number    = {12},
  pages     = {1929--1941},
  year      = {2011},
  publisher = {Nature Publishing Group}
}

@article{foi2008practical,
  title     = {Practical {Poissonian-Gaussian} noise modeling and fitting for single-image raw-data},
  author    = {Foi, Alessandro and Trimeche, Mejdi and Katkovnik, Vladimir and Egiazarian, Karen},
  journal   = {IEEE Transactions on Image Processing},
  volume    = {17},
  number    = {10},
  pages     = {1737--1754},
  year      = {2008},
  publisher = {IEEE}
}

@article{flamary2021pot,
  title   = {{POT}: {Python} optimal transport},
  author  = {Flamary, R{\'e}mi and Courty, Nicolas and Gramfort, Alexandre and Alaya, Mokhtar Z and Boisbunon, Aur{\'e}lie and Chambon, Stanislas and Chapel, Laetitia and Corenflos, Adrien and Fatras, Kilian and Fournier, Nemo and others},
  journal = {Journal of Machine Learning Research},
  volume  = {22},
  number  = {78},
  pages   = {1--8},
  year    = {2021}
}

@inproceedings{li2023lsdir,
  title     = {{LSDIR}: A large scale dataset for image restoration},
  author    = {Li, Yawei and Zhang, Kai and Liang, Jingyun and Cao, Jiezhang and Liu, Ce and Gong, Rui and Zhang, Yulun and Tang, Hao and Liu, Yun and Demandolx, Denis and others},
  booktitle = {Proceedings of the IEEE/CVF Conference on Computer Vision and Pattern Recognition (CVPR)},
  pages     = {1775--1787},
  year      = {2023}
}

@inproceedings{tan2019efficientnet,
  author    = {Tan, Mingxing and Le, Quoc V.},
  title     = {{EfficientNet}: Rethinking model scaling for convolutional neural networks},
  booktitle = {Proceedings of the International Conference on Machine Learning (ICML)},
  series    = {Proceedings of Machine Learning Research},
  volume    = {97},
  pages     = {6105--6114},
  year      = {2019}
}

@inproceedings{lim2017enhanced,
  author    = {Lim, Bee and Son, Sanghyun and Kim, Heewon and Nah, Seungjun and Lee, Kyoung Mu},
  title     = {Enhanced deep residual networks for single image super-resolution},
  booktitle = {Proceedings of the IEEE Conference on Computer Vision and Pattern Recognition (CVPR) Workshops},
  pages     = {136--144},
  year      = {2017}
}

@inproceedings{Agustsson_2017_CVPR_Workshops,
  author    = {Agustsson, Eirikur and Timofte, Radu},
  title     = {{NTIRE} 2017 challenge on single image super-resolution: Dataset and study},
  booktitle = {Proceedings of the IEEE Conference on Computer Vision and Pattern Recognition (CVPR) Workshops},
  month     = {July},
  year      = {2017}
}

@article{carbajal2023blind,
  title     = {Blind motion deblurring with pixel-wise kernel estimation via kernel prediction networks},
  author    = {Carbajal, Guillermo and Vitoria, Patricia and Lezama, Jos{\'e} and Mus{\'e}, Pablo},
  journal   = {IEEE Transactions on Computational Imaging},
  volume    = {9},
  pages     = {928--943},
  year      = {2023},
  publisher = {IEEE}
}

@article{ipol.2019.243,
  title   = {Blind image deblurring using the {$\ell_0$} gradient prior},
  author  = {Anger, J{\'e}r{\'e}my and Facciolo, Gabriele and Delbracio, Mauricio},
  journal = {Image Processing On Line},
  volume  = {9},
  pages   = {124--142},
  year    = {2019},
  doi     = {10.5201/ipol.2019.243}
}

@inproceedings{pan2014deblurring,
  title     = {Deblurring text images via {$L_0$}-regularized intensity and gradient prior},
  author    = {Pan, Jinshan and Hu, Zhe and Su, Zhixun and Yang, Ming-Hsuan},
  booktitle = {Proceedings of the IEEE Conference on Computer Vision and Pattern Recognition (CVPR)},
  pages     = {2901--2908},
  year      = {2014}
}

@inproceedings{ren2020neural,
  title     = {Neural blind deconvolution using deep priors},
  author    = {Ren, Dongwei and Zhang, Kai and Wang, Qilong and Hu, Qinghua and Zuo, Wangmeng},
  booktitle = {Proceedings of the IEEE/CVF Conference on Computer Vision and Pattern Recognition (CVPR)},
  pages     = {3341--3350},
  year      = {2020}
}

@article{qiao2024deep,
  title     = {Deep learning-based optical aberration estimation enables offline
               digital adaptive optics and super-resolution imaging},
  author    = {Qiao, Chang and Chen, Haoyu and Wang, Run and Jiang, Tao and
               Wang, Yuwang and Li, Dong},
  journal   = {Photonics Research},
  volume    = {12},
  number    = {3},
  pages     = {474--484},
  year      = {2024},
  publisher = {Optica Publishing Group},
  doi       = {10.1364/PRJ.506778}
}

@article{qiao2021biosr,
  title   = {Evaluation and development of deep neural networks for image super-resolution in optical microscopy},
  author  = {Qiao, Chang and Li, Di and Guo, Yuting and Liu, Chong and Jiang, Tao and Dai, Qionghai and Li, Dong},
  journal = {Nature Methods},
  volume  = {18},
  pages   = {194--202},
  year    = {2021},
  doi     = {10.1038/s41592-020-01048-5}
}

@article{guo2025deep,
  title     = {Deep learning-based aberration compensation improves contrast and
               resolution in fluorescence microscopy},
  author    = {Guo, Min and Wu, Yicong and Hobson, Chad M. and Su, Yijun and
               Qian, Shuhao and Krueger, Eric and Christensen, Ryan and
               Kroeschell, Grant and Bui, Johnny and Chaw, Matthew and
               Zhang, Lixia and Liu, Jiamin and Hou, Xuekai and Han, Xiaofei and
               Lu, Zhiye and Ma, Xuefei and Zhovmer, Alexander S. and
               Combs, Christian and Moyle, Mark and Yemini, Eviatar and
               Liu, Huafeng and Liu, Zhiyi and Benedetto, Alexandre and
               La Riviere, Patrick and Col{\'o}n-Ramos, Daniel A. and Shroff, Hari},
  journal   = {Nature Communications},
  volume    = {16},
  number    = {1},
  pages     = {313},
  year      = {2025},
  publisher = {Nature Publishing Group},
  doi       = {10.1038/s41467-024-55267-x}
}

@article{ji2017adaptive,
  title     = {Adaptive optical fluorescence microscopy},
  author    = {Ji, Na},
  journal   = {Nature Methods},
  volume    = {14},
  number    = {4},
  pages     = {374--380},
  year      = {2017},
  publisher = {Nature Publishing Group},
  doi       = {10.1038/nmeth.4218}
}

@article{koho2020two,
  title   = {Two-photon image-scanning microscopy with {SPAD} array and blind image reconstruction},
  author  = {Koho, Sami V. and Slenders, Eli and Tortarolo, Giorgio and Castello, Marco and Buttafava, Mauro and Villa, Federica and Tcarenkova, Elena and Ameloot, Marcel and Bianchini, Paolo and Sheppard, Colin J. R. and Diaspro, Alberto and Tosi, Alberto and Vicidomini, Giuseppe},
  journal = {Biomedical Optics Express},
  volume  = {11},
  number  = {6},
  pages   = {2905--2924},
  year    = {2020},
  doi     = {10.1364/BOE.374398}
}

@article{premillieu2024aberration,
  title   = {Aberration correction in epi-fluorescence microscopy using unknown speckle illumination},
  author  = {Premillieu, Evolene and Caravaca-Aguirre, Antonio M. and Labouesse, Simon and Irsch, Kristina and Piestun, Rafael},
  journal = {Optics Express},
  volume  = {32},
  number  = {21},
  pages   = {37658--37667},
  year    = {2024},
  doi     = {10.1364/OE.526148}
}

@article{matsuda2026phase,
  title   = {Phase-based computational adaptive optics enables artifact-free super-resolution microscopy},
  author  = {Matsuda, Atsushi and Rodriguez-Reza, Carlos Mario and Tamada, Yosuke and Matsuo, Yamato and Yamamoto, Takaharu G. and Koujin, Takako and Carlton, Peter M.},
  journal = {Communications Engineering},
  volume  = {5},
  pages   = {75},
  year    = {2026},
  doi     = {10.1038/s44172-026-00622-7}
}

@misc{qdo2026phase,
  title  = {Diversity calibration for {PSF} estimation},
  author = {Do, Quoc-Bao and Nguyen, Minh-Hai and Soubies, Emmanuel and Weiss, Pierre},
  note   = {Preprint},
  year   = {2026}
}

@article{stringer2025cellpose3,
  title   = {{Cellpose3}: one-click image restoration for improved cellular segmentation},
  author  = {Stringer, Carsen and Pachitariu, Marius},
  journal = {Nature Methods},
  volume  = {22},
  number  = {3},
  pages   = {592--599},
  year    = {2025},
  doi     = {10.1038/s41592-025-02595-5}
}

@article{stringer2021cellpose,
  title   = {{Cellpose}: a generalist algorithm for cellular segmentation},
  author  = {Stringer, Carsen and Wang, Tim and Michaelos, Michalis and Pachitariu, Marius},
  journal = {Nature Methods},
  volume  = {18},
  pages   = {100--106},
  year    = {2021},
  doi     = {10.1038/s41592-020-01018-x}
}

@article{mazzella2024extended,
  title   = {Extended-depth of field random illumination microscopy, {EDF-RIM}, provides super-resolved projective imaging},
  author  = {Mazzella, Lorry and Mangeat, Thomas and Giroussens, Guillaume and Rogez, Benoit and Li, Hao and Creff, Justine and Saadaoui, Mehdi and Martins, Carla and Bouzignac, Ronan and Labouesse, Simon and Idier, J{\'e}r{\^o}me and Galland, Fr{\'e}d{\'e}ric and Allain, Marc and Sentenac, Anne and LeGoff, Lo{\"i}c},
  journal = {Light: Science \& Applications},
  volume  = {13},
  number  = {1},
  pages   = {285},
  year    = {2024},
  doi     = {10.1038/s41377-024-01612-0}
}

@article{ainslie2025desmoplakin,
  title   = {The desmoplakin tail domain position in the desmosomal plaque is isoform dependent},
  author  = {Ainslie, Collin M. and Patel, Krishna and Tran, Yen T. B. and Bartley, Samuel C. and Bharathan, Navaneetha Krishnan and Spindler, Volker and Mattheyses, Alexa L.},
  journal = {Journal of Cell Science},
  volume  = {138},
  number  = {17},
  pages   = {jcs263906},
  year    = {2025},
  doi     = {10.1242/jcs.263906}
}

@article{faklaris2022quality,
  title   = {Quality assessment in light microscopy for routine use through simple tools and robust metrics},
  author  = {Faklaris, Orestis and Bancel-Vall{\'e}e, Leslie and Dauphin, Aur{\'e}lien and Monterroso, Baptiste and Fr{\`e}re, Perrine and Geny, David and Manoliu, Tudor and de Rossi, Sylvain and Cordeli{\`e}res, Fabrice P. and Schapman, Damien and Nitschke, Roland and Cau, Julien and Guilbert, Thomas},
  journal = {Journal of Cell Biology},
  volume  = {221},
  number  = {11},
  pages   = {e202107093},
  year    = {2022},
  doi     = {10.1083/jcb.202107093}
}

\vspace{4pt}
\noindent\textbf{Acknowledgements}\\
{\fontsize{8.5}{10}\selectfont
    The authors acknowledge support from the ANR Micro-Blind ANR-21-CE48-0008 and from the ANR CLEAR Microscopy ANR-25-CE45-3780.
    {This work drew on the computing resources of the Jean Zay supercomputer, funded by the French National High-Performance Computing
        Infrastructure (GENCI) and operated by the CNRS Institute for Development and Resources in Scientific Computing (IDRIS), as well as support from
    IDRIS teams (GENCI-IDRIS allocation AD011012210).}

    The authors thank Emmanuel Soubies and Quoc Bao Do for fruitful discussions and for their help in designing the PSF estimation algorithm from the
    sharp/aberrated image pairs.

    {The authors thank N\'emo Bouillon and Alexandre Huyghe, engineers in the CNRS Programme national de recherche en intelligence artificielle
        (PNRIA) hosted at IRIT, for their help in designing the neural network architecture and training procedure. This work benefited from the
        support of engineers in
        artificial intelligence within a program funded by the French government through the France 2030 investment plan managed by the National
    Research Agency (ANR), under reference ANR-24-RIIA-0001.}

}

\vspace{4pt}
\noindent\textbf{Author contributions}\\
{\fontsize{8.5}{10}\selectfont
    M.H.N. and F.S. implemented the method, performed the experiments, and wrote the paper.
    T.M. designed the deformable-mirror experiment and the RIM experiments.
    P.E. designed the method and designed the noise level identification module.
    S.C. performed the deformable-mirror experiments.
    P.W. supervised the project, designed the method, implemented the training pipeline, and wrote the paper.
    All authors reviewed the manuscript.
}

\vspace{4pt}
\noindent\textbf{Competing interests}\\
{\fontsize{8.5}{10}\selectfont
    The authors declare no competing interests.
}

\vspace{4pt}
\noindent\textbf{Data availability}\\
{\fontsize{8.5}{10}\selectfont
    The deformable-mirror dataset comprises 1{,}332 acquisitions (1{,}290 aberrated images and 42 flat-mirror images) and is available
    at \url{https://doi.org/10.5281/zenodo.24304044}.
}

\vspace{4pt}
\noindent\textbf{Code availability}\\
{\fontsize{8.5}{10}\selectfont
    The backbone of the method is implemented in PyTorch and is available within the deepinv package
    \url{https://deepinv.org/auto_examples/physics/demo_blur_tour.html}.
    A demo is available at \url{https://deepblur.mambo-utoulouse.fr}.
    The Napari plugin is available at \url{https://pypi.org/project/deepblur-napari/}.
}

\clearpage
\onlinemethods

\subsection{Microscope simulation}\label{sec:image_formation}

To create a realistic database of pairs of images and PSFs, we synthesize an observation \(y\) from a clean image \(x\) with normalized values in
\([0,1]\) as follows:
\begin{equation}\label{eq:forward_model}
    y = \mathcal{Q}_\mathrm{b} \left( \mathcal{P}_{\gamma}(A_w (I \odot  x)) + n_\sigma \right)
\end{equation}
where \(\mathcal{P}_\gamma(z)\) draws Poisson counts of mean \(\gamma z\), so that the gain \(\gamma\) is the expected photon count of a
unit-intensity pixel, \(n_\sigma \sim \mathcal{N}(0, \sigma^2 \mathrm{Id})\) is Gaussian read noise expressed in photon
counts~\cite{foi2008practical}, and \(\mathcal{Q}_\mathrm{b}\) rounds and clips the result to \(\mathrm{b}\)-bit integers.
The degradation operator that maps a sharp image to its blurry version is denoted by \(A_w\), where \(w\) represents the parameters of the
degradation (the Zernike coefficients).
The excitation pattern is denoted \(I\), with \(\odot\) being the element-wise multiplication.
This model covers both standard widefield microscopy with \(I\) being a near constant and smoothly varying illumination and SIM or RIM
microscopy~\cite{Mangeat2021} where \(I\) varies at higher frequencies. The parameters of the model are drawn randomly at training time, to ensure
that the network learns to be robust to a wide range of imaging conditions. The ranges and distributions of these parameters are summarized
in~Extended Data Table~\ref{tab:simulation_parameters}.
\paragraph{PSF model}

We use a scalar diffraction model of an incoherent microscope with a circular aperture: the intensity PSF \(h_w\) is defined as the squared magnitude
of the Fourier transform \(\mathcal{F}\) of the pupil function \(P_w\) (Debye approximation):
\begin{equation}
    h_w = \left| \mathcal{F} \left( P_w(\rho, \theta) \right) \right|^2,
\end{equation}
with
\begin{equation}
    P_w(\rho, \theta) = A(\rho)\exp \left(- i 2 \pi \Phi_w(\rho, \theta)\right),
\end{equation}
where \(A\) is a binary circular aperture function, and \((\rho, \theta)\) are normalized
polar coordinates in the pupil plane. The coherent cutoff frequency \(f_c = \mathrm{NA}\cdot p/\lambda\), in cycles per pixel
(\(p\): pixel size),
corresponds to \(\rho = 1\), so the optical transfer function (OTF) is band-limited at \(2f_c\). The wavefront error
\(\Phi_w\), expressed in waves, is expanded onto the OSA/ANSI Zernike basis with Noll (unit-RMS) normalization:
\begin{equation}
    \Phi_w(\rho, \theta) = \sum_{k \in K} w_k Z_k(\rho, \theta),
\end{equation}
where \(w = (w_k)_{k \in K} \in \mathbb{R}^{|K|}\) are the Zernike coefficients~\cite{Lakshminarayanan10042011,noll1976zernike}.
Zernike's polynomials form an orthogonal basis over the circular pupil and map directly onto
classical aberrations (astigmatism, coma, spherical aberration), which makes the simulated aberrations physically
interpretable.

\paragraph{Multi-spectral PSFs}
Each colour channel \(c=1,\dots,C\) is assigned an emission wavelength \(\lambda_c\), drawn uniformly in \([400, 700]\)\,nm and
sorted so that \(\lambda_1\) is the shortest. As the numerical aperture and the pixel size are common to all channels, the cutoff
frequency scales as
\begin{equation}
    f_c(\lambda_c) = f_c(\lambda_1)\,\frac{\lambda_1}{\lambda_c},
\end{equation}
with \(f_c(\lambda_1)\) drawn as in~Extended Data Table~\ref{tab:simulation_parameters}. For non-dispersive aberrations, the optical path difference does
not depend on the wavelength, so the coefficients, expressed in waves, scale as
\begin{equation}
    w_k(\lambda_c) = w_k(\lambda_1)\,\frac{\lambda_1}{\lambda_c} + \varepsilon_{k,c},
    \quad \varepsilon_{k,c} \sim \mathcal{N}(0, 0.005^2),
\end{equation}
where \(w_k(\lambda_1)\) follows~\eqref{eq:zernike_coeff_dist} and \(\varepsilon_{k,c}\) models weak dispersion. Each training batch
uses one to three channels, randomly selected among the RGB channels of the source images.

\paragraph{Discretization and sampling}
The simulator is implemented in PyTorch on a discrete Fourier grid within the DeepInverse package~\cite{Tachella2025}. The pupil is
sampled on a \(1024 \times 1024\) grid spanning normalized frequencies \([-0.5, 0.5]\), and PSFs are cropped to \(71 \times 71\) pixels.
The upper bound \(f_c \leq 0.25\) ensures that the incoherent cutoff \(2f_c\) respects the Nyquist limit; the lower bound
\(f_c = 0.125\) corresponds to twofold oversampling.
\paragraph{PSF distribution}
We use all Zernike modes of radial order \(2 \leq n \leq 7\), i.e.\ \(Z_3\)--\(Z_{35}\) in the OSA/ANSI indexing (33 modes).
Piston, tip, and tilt (\(n=0,1\)) are excluded as they do not affect the intensity PSF;
defocus (\(Z_4\)) is included and accounts for out-of-focus blur within the paraxial
approximation. The coefficients are sampled with an amplitude that decreases with the radial order:
\begin{equation}\label{eq:zernike_coeff_dist}
    \begin{gathered}
        w_k \sim \mathcal{U}(-a_k, a_k), \quad a_k = \frac{a}{2}\min\left(1, \frac{9}{n_k^2}\right), \\
        a = 0.1 + 0.2\,u^{1/3}, \quad u \sim \mathcal{U}(0, 1),
    \end{gathered}
\end{equation}
where \(n_k\) is the radial order of mode \(k\) and \(a\) controls the overall aberration
level, biased toward strong aberrations by the cube root. Modes up to radial order 3 share the same range, while
higher-order modes, which have a smaller perceptual impact, are attenuated as \(n_k^{-2}\). Sample PSFs are shown in Extended Data
Fig.~\ref{edfig:example_PSFs}.

\paragraph{Spatially varying PSF}
The network is trained on spatially invariant PSFs; spatially varying PSFs are simulated only for synthetic tests.
We compute PSFs on a \(5 \times 5\) grid of overlapping \(126 \times 126\) patches (stride 50 pixels) covering a \(326 \times 326\) field
of view, and interpolate them via a product-convolution expansion~\cite{Escande2017}.
Smooth variations are obtained from a spatially correlated random field \(\xi = (\xi_p)_{1 \leq p \leq 25}\), \(\xi_p \in \mathbb{R}^{|K|}\),
computed by smoothing white Gaussian noise on the patch grid with a Gaussian kernel of width drawn uniformly in \([2, 8]\) grid units,
and normalized so that \(\max_p |\xi_{p,k}| = 1\) for each mode \(k\).
The local coefficients perturb a base vector \(\bar w\) drawn from~\eqref{eq:zernike_coeff_dist}:
\begin{equation}
    w_{p,k} = \mathrm{clip}_{[-a_k, a_k]}\left(\bar w_k + 2 s\, a_k\, \xi_{p,k}\right), \quad s \sim \mathcal{U}(0, 0.5),
\end{equation}
where \(s\) sets the strength of the spatial variations. The resulting operator \(A_w\) is parametrized by
\(w = (w_p)_{1 \leq p \leq 25}\).
Examples are shown in Extended Data Fig.~\ref{edfig:example_PSFs}.

On real and synthetic data with spatially varying blurs, since the network is trained on spatially invariant blur, we estimate the PSF of each input
patch, which we assign to the patch
centre. Applying it to overlapping tiles (\(256 \times 256\) pixels, stride 128) yields a map of local PSFs across the field of view,
which the product-convolution expansion~\cite{Escande2017} turns into a space-varying blur operator for computational imaging.

\paragraph{Excitation pattern}

For grayscale images, the training sets of both \name and RAM use a uniform excitation pattern \(I\) for two thirds of the
samples and a speckle pattern for the remaining third; the chromatic model is trained with uniform illumination only.
The speckle illumination patterns are dedicated to the RIM experiments and are obtained by filtering a complex Gaussian white noise through the
system pupil function \(P_w\).
The illumination pattern is given by the squared magnitude of the speckle field \(I = |E|^2\), where
\begin{equation}
    E = \mathcal{F}^{-1} \left[ P_w \cdot \eta \right],
    \quad \text{with} \quad \eta \sim \mathcal{CN}(0,1).
\end{equation}

\subsection{PSF estimation network}
Previous works~\cite{debarnot2024deep,shajkofci2020spatially,Jin:18} have employed standard backbones such as ResNet, ResNeXt or AlexNet, which are
originally designed for image classification tasks, to regress Zernike coefficients.
These architectures may not be optimal for PSF estimation in the pixel space.
In particular, we have found that batch normalization has a detrimental effect on the performance of PSF estimation
networks, with a drop in accuracy and stability on out-of-distribution data.
In this work, we propose a deep convolutional neural network whose architecture is designed to capture both the high-frequency diffraction rings and
the global structure of the blur kernel, specifically tailored
for PSF estimation in pixel space, see Extended Data Fig.~\ref{edfig:architecture} for an illustration of the architecture design.

The architecture comprises three components, a residual feature encoder, a multi-scale context aggregation module and a PSF
regression head, and uses no batch normalization.
\begin{itemize}
    \item \textbf{Feature encoder.} The backbone of our model is a residual network inspired by ResNet~\cite{he2016deep}. The encoder consists of an
        initial convolutional layer followed by five stages of Residual Blocks.
        We progressively increase the channel width from \(64\) to \(1024\) (64, 128, 256, 512, 768, 1024) to preserve rich feature
        representations of the microscopic structures, using
        kernel sizes of \(3 \times 3\) and a \(1 \times 1\) convolution for the skip connections.
        A key design choice is the use of \emph{Instance Normalization (IN)}~\cite{ulyanov2016instance} throughout the encoder blocks, which were initially
        designed for style transfer tasks.
        In microscopy, image intensity can fluctuate significantly due to varying exposure times or fluorophore concentrations. IN renders the feature
        extraction invariant to these global contrast and illumination shifts, reducing sensitivity to feature-wise intensity scale and offset. We
        empirically found that using IN in early stages significantly improves the robustness of
        PSF estimation compared to Batch Normalization or Group Normalization.
        We use GELU activations~\cite{hendrycks2016gaussian} throughout, which we found to yield slightly better convergence than ReLU in preliminary
        experiments.
    \item \textbf{Multi-scale context aggregation.}
        Diffraction patterns in microscopy can vary significantly in scale depending on the numerical aperture and defocus distance. To capture these
        multi-scale dependencies, we incorporate an Atrous Spatial Pyramid Pooling (ASPP) module~\cite{chen2017rethinking} at the bottleneck of the encoder.
        The ASPP module combines a \(1 \times 1\) convolution and three \(3 \times 3\) dilated convolutions with rates
        \(r \in \{2, 4, 8\}\), whose outputs are concatenated and projected back by a \(1 \times 1\) convolution; like the decoder, it
        uses Group Normalization. This allows the network to effectively enlarge its
        receptive field and integrate context from the entire diffraction pattern without downsampling the spatial resolution further.
    \item \textbf{PSF regression head.}
        The decoder transforms the aggregated features into the target PSF of size \(P \times P\). Following two decoding blocks that reduce the channel
        dimension, we employ a projection head. A final convolutional layer maps the features to \(P^2\) channels with a spatial resolution of
        \(\frac{H}{32} \times \frac{W}{32}\), which is resampled to \(8 \times 8\) by bilinear interpolation (an identity for
        \(256 \times 256\) inputs).
        To recover the spatial structure of the PSF, we perform a dimension rearrangement.
        Specifically, the \(P^2\) output channels are reshaped into a \(P\times P\) spatial grid, with the \(8\times 8\) spatial dimensions acting as latent
        descriptors that are linearly mixed via a \(1\times1\) convolution to produce the final PSF values.
        The learnable temperature \(\tau\) controls the sharpness of the output distribution, allowing the network to adapt between peaked and diffuse PSF
        shapes during training:
        \begin{equation}
            \hat{h} = \mathrm{softmax} \left( \frac{\mathrm{output}}{\tau} \right).
        \end{equation}
        While the encoder prioritizes invariance, the regression head requires stable feature aggregation to reconstruct the precise kernel values.
        Therefore, we switch to \emph{Group Normalization (GN)}~\cite{wu2018group} in the decoding and regression layers. GN provides a stable optimization
        landscape for the regression task without the batch-dependency of Batch Normalization, which is particularly advantageous given the high dynamic
        range of the inferred PSF features.
    \item \textbf{Multi-channel input and output.} Images with one, two or three channels are mapped by a channel-count-specific
        \(3 \times 3\) convolution to a common 4-channel stem, so that the rest of the network is shared. The final \(1 \times 1\) mixing
        convolution outputs one PSF per input channel; all other layers are shared across channels, since the channel PSFs are
        correlated deformations of the same optical system.
\end{itemize}
We let \(N_\theta\) denote the neural network with parameters \(\theta\), and the estimated PSF is denoted by \(\hat{h}(y) = N_\theta(y)\).

\paragraph{Receptive field analysis.}
Given an input resolution of \(256 \times 256\) pixels, our encoder reduces the spatial dimensionality by a factor of \(16\), resulting in a \(16
\times 16\) bottleneck tensor.
To ensure the network captures global diffraction patterns, the receptive field ideally encompasses the entire field of view.
While the standard residual encoder provides a maximum receptive field of \(157\) pixels, the inclusion of the ASPP module significantly expands this
range and combines multiscale context.
The branch with the largest dilation rate (\(r=8\)) operates on the bottleneck features with an effective kernel span of \(17 \times 17\) spatial
units, which is larger than the \(16 \times 16\) spatial extent of the bottleneck itself.
This ensures that the estimated PSF at any spatial location is informed by the global context of the input image.

\paragraph{Computational complexity.}
Despite its relatively large size with \(98\)M parameters, the model runs inference in approximately 8\,ms per \(256 \times 256\) patch on a
workstation GPU (NVIDIA RTX 4000
Ada) and about 80\,ms on CPU (Intel Xeon W-2275).

\paragraph{Training}
The network is trained by minimizing the \(\ell^1\) distance \(\|\hat h - h\|_1\) between estimated and ground-truth PSFs, averaged
over the batch and the PSF channels. Clean images are random crops from ImageNet~\cite{deng2009imagenet},
Flickr2K~\cite{lim2017enhanced}, DIV2K~\cite{Agustsson_2017_CVPR_Workshops} and LSDIR~\cite{li2023lsdir}. To mimic fluorescence images, a random
subset of them is converted into sparse structures, by soft-thresholding at a
random intensity quantile between 0.7 and 0.99, into thresholded Sobel edge maps, or into a mixture of both (\Cref{fig:method}b). Each
simulated observation is normalized by clipping its intensities to its 1st and 99th percentiles and rescaling them to \([0, 1]\). We
use Adam~\cite{kingma2017adammethodstochasticoptimization} with a learning rate of \(10^{-4}\) after a linear warm-up of 1,000 steps, a
batch size of 64. We train for 50 epochs on blurry patches of size \(256 \times 256\), then adapt the model for \(10\) additional epochs on blurry
patches of random height and width between 128 and 256 pixels. Since PSFs, noise, illumination and augmentations are drawn on the fly, training pairs are
effectively never repeated, which limits the risk of overfitting despite the 98 million parameters of the network.

\subsection{Noise level estimation}

A separate network estimates the parameters \((\gamma, \sigma)\) of the Poisson--Gaussian noise model~\eqref{eq:forward_model},
which are required by RAM. It is based on EfficientNet-B0~\cite{tan2019efficientnet}, trained from scratch, with a single-channel
input layer, all normalization layers replaced by instance normalization, and a three-layer perceptron head
(\(1000 \to 512 \to 32 \to 2\) units) whose final ReLU ensures non-negative outputs; it has 5.8 million parameters. Since the input
image is rescaled to \([0, 1]\) by min--max normalization, its absolute intensity is lost, and the gain must be inferred from the
signal dependence of the noise. The network is trained with the simulator described above, with \(\gamma \in [2^5, 2^{12}]\) and
\(\sigma \in [0.5, 5]\) photon counts (both log-uniform), Zernike PSFs up to radial order 4 and uniform illumination, by minimizing
\begin{equation}
    \frac{|\hat\sigma - \sigma|}{\sigma_{\max}} + \frac{|\hat\gamma - \gamma|}{\gamma_{\max}},
\end{equation}
averaged over the batch, with \(\sigma_{\max} = 5\) and \(\gamma_{\max} = 2^{12}\). We use Adam with a learning rate of \(10^{-4}\) for 90 epochs
after 1,000 warm-up steps, a batch size of 128 and an exponential moving average of the weights (decay 0.999). For colour images,
each channel is processed independently, and the estimates are passed to RAM as \(\gamma^{-1}\) and \(\sigma/\gamma\).

\subsection{RAM deconvolution model}

For non-blind deconvolution, we use the Reconstruct Anything Model (RAM)~\cite{terris2026reconstruct}, available in
DeepInverse~\cite{Tachella2025}. RAM is a lightweight convolutional network that receives the forward operator \(A\), through
\(A^\top y\) and physics-informed blocks, together with the parameters of a Poisson--Gaussian noise model, so that a single set of
weights applies to arbitrary linear operators. We keep its architecture unchanged and fine-tune all its weights, starting from the
pretrained model, on data generated by our microscope simulator~\eqref{eq:forward_model}.

The fine-tuning data are generated as for \name: same clean images and fluorescence-like augmentations, Zernike PSFs drawn
from~\eqref{eq:zernike_coeff_dist} with a maximal radial order drawn at random for each batch, Poisson--Gaussian noise, and speckle
illumination for one third of the grayscale samples. A colour version is fine-tuned in the same way with the multi-spectral PSFs
described above. RAM is conditioned on the noise level through the Poisson gain and Gaussian standard deviation expressed in the
intensity scale of the clean image, \(\gamma^{-1}\) and \(\sigma/\gamma\). During fine-tuning, RAM receives the space-invariant true PSF; at
inference, RAM receives the space-varying product-convolution operator built from the \name estimates.
Training with the true PSF ensures that the network learns to invert the physics of the microscope rather than to compensate for errors in the PSF
estimate; it also reduces the computational cost of training.

To avoid boundary artefacts, the forward operator is a valid convolution: the clean image extends beyond the observed field by half
the PSF width on each side, RAM reconstructs this extended field, and the loss is computed on the central region matching the
observation. The loss is the \(\ell^1\) error between the reconstruction and the clean image, weighted for each sample by
\(\gamma \|A^\top y\|_2\). We use Adam~\cite{kingma2017adammethodstochasticoptimization} with a learning rate of \(10^{-4}\) after a
linear warm-up of 1,000 steps, and keep an exponential moving average of the weights (decay 0.999).
The network was fine tuned for 60 000 iterations with a batch size of 32 on \(256 \times 256\) images.

\subsection{Deformable mirror experimental setup}

We used a Nikon-based widefield microscope equipped with a deformable mirror (DM) to introduce controlled aberrations.
The DM allows us to apply known wavefront distortions, enabling quantitative validation of our PSF estimation method under realistic conditions.
The optical setup is shown in Fig.~\ref{fig:dm}a.

\paragraph{Widefield with commercial adaptive optics system}
Adaptive optics experiments were performed using a Nikon Ti-E/B automated inverted epifluorescence microscope. This was
equipped with a `Perfect Focus System' (PFS, Nikon), a piezo stage (Nano Z100-N, Mad City Labs), a CFI Apochromat TIRF 100XC oil
objective (NA 1.49), a quad-band dichroic mirror (97335, Nikon N-STORM TIRF Filter Set), a widefield illumination arm, an ORCA-Flash4.0
V3 sCMOS camera (C13440-20C, Hamamatsu), and a thermostatic chamber at 24\,\(^\circ\)C to limit instrumental drift. Excitation laser power is controlled
by an AOTF. A 488\,nm laser (Coherent) was used to excite Argolight or podosomes, and a 561\,nm laser (Coherent Sapphire) was used for
imaging at a power of 1\,kW\,cm\(^{-2}\). The collection point spread function (PSF) was deformed using adaptive optics with a
deformable mirror (MicAO 3DSR, Imagine Optics) inserted in the detection path just before the sCMOS camera. The mirror is placed on
the Fourier plane and the total magnification of the collection system is around 1; the effective pixel size in the specimen plane was
65\,nm.

Although the mirror control software allows us to directly specify the Zernike coefficients applied to the DM, we do not use this information for
PSF estimation.
Indeed, the DM calibration is not perfect, and we observed that the actual wavefront deviates significantly from the commanded settings due to factors
such as hysteresis, nonlinearity, or mechanical imperfections.
We designed a calibration algorithm reminiscent of phase diversity methods to estimate the actual PSF corresponding to each DM setting.
It serves as the ground truth for evaluating our method.
The idea is to acquire one image \(y_0\) with the DM in non-aberrated state and another one \(y_1\) with an aberration.
Hence, we can model the images as
\begin{align*}
    y_0 & = x \star h_0 + e_0  \\
    y_1 & = x \star h_1 + e_1,
\end{align*}
where \(x\) is an unknown image, \(h_0\) is an Airy pattern and \(h_1\) is the PSF we would like to recover.
Various approaches can be used to estimate \(h_1\) from \(y_0\) and \(y_1\).
For instance, we can minimize \(\|h\star y_0 - h_0 \star y_1\|^2\) over the set of physically realistic PSFs.
In the preprint \cite{qdo2026phase}, we designed a specific neural network for this problem and showed that it provides the most reliable estimate of the PSF.

By comparing the estimated PSF with the reference PSF derived, we can directly assess the accuracy of our approach in recovering complex aberration patterns.
For this experiment, we generated a dataset of \(31\) different aberrations by applying various combinations of Zernike modes with varying
amplitudes, covering a wide range of realistic optical distortions. Reference PSFs for representative mirror states, for one field
of view of each sample, are shown in Extended Data Fig.~\ref{edfig:estimated_PSFs_DM}.

\subsection{Deformable-mirror quantitative evaluation}

For the per-image distributions in Extended Data Fig.~\ref{edfig:dm_violins}, we analysed the 1{,}290 aberrated images and excluded the 42
acquisitions with a flat mirror. PSF estimates were compared with the independently estimated reference PSFs using the metrics defined below.
Each method's PSF was supplied to the same fine-tuned RAM model. For each image-quality metric \(M\), we computed the paired reconstruction gain
\(\Delta M=M(\hat{x},x)-M(y,x)\), where \(y\) is the aberrated input, \(\hat{x}\) the restoration and \(x\) the reference image. Positive gains
indicate improvement for PSNR and SSIM; negative gains indicate improvement for LPIPS. Violin plots show the distributions, medians and quartiles.

\subsection{Random illumination microscopy}

\paragraph{RIM system for imaging desmoplakin}
We use an inverted microscope (TEi Nikon) with a 100\(\times\) TIRF objective (CFI SR APO 100\(\times\) NA 1.49, Nikon).
For GFP/Alexa-488 imaging, we use a 488\,nm diode laser (Oxxius LBX-488-200-CSB). A fast spatial light phase binary modulator (SLM, QXGA Fourth
Dimension), which was placed in an image plane, was illuminated by the collimated 8.8\,mm TEM00 beam. After passing through the dichroic mirror
(Semrock, DI02-R488-25\(\times\)36) and Stop-Line notch filter (Semrock, NF03-405/488/561/635E-25), fluorescence is collected by an sCMOS camera
(OrcaFlash Fusion). A stair-step phase mask (DoubleHelix, PNDT6) was placed in the pupil plane of an optical relay consisting of \(f_{R1}\) and
\(f_{R2}\) (with focal lengths of 200\,mm and 300\,mm, respectively). A bandpass filter (Semrock FF01-514/30-25) was used for green fluorescent
protein. The INSCOPER software and controller provide the user interface and synchronisation of all instruments.

\paragraph{RIM system for imaging \textit{Drosophila} egg chambers}
3D images were acquired using an inverted microscope (TEi Nikon) equipped with a \(\times\)60 magnification, 1.35 NA CFI Plan Apochromat Lambda S
60XC Sil Nikon objective and an ORCA-Quest sCMOS camera (Hamamatsu). Commercial acquisition software (INSCOPER SA) enabled whole-cell,
single-timepoint 3D-RIM acquisition in only 6\,s under a low-photobleaching regime (1\,W\,cm\(^{-2}\)). Fast diode lasers (Oxxius) with wavelengths
centred at 488\,nm and 561\,nm (LBX-488-200-CSB and LBX-561-200-CSB, respectively) were used to produce a TEM00 beam with a diameter of 2.2\,mm. The
polarised beam was rotated by 5\(^\circ\) before passing through a \(\times\)4 beam expander (GBE04-A), producing an 8.8\,mm TEM00 beam. A fast
spatial light phase binary modulator (QXGA, Fourth Dimensions) was conjugated to the image plane to create 200 RIMs per plane~\cite{Mangeat2021}.

\paragraph{RIM image reconstruction}
For each spatial region, we estimated independently a collection PSF with \name{} from the Widefield image --
obtained as the sum of all speckle-illuminated frames.
The reconstruction is essentially based on the algoRIM treatment~\cite{Mangeat2021} developed for the analysis of 2D-RIM
images, and which is accessible at \url{https://github.com/teamRIM/tutoRIM}. The \(N\) images taken under different speckled
illumination (\(N \sim 100\)--200 in most cases) are first processed with a Wiener filter, which can be written in the Fourier domain:
\begin{equation*}
    \tilde{g} = \frac{\tilde{h}_\perp^{*}}{|\tilde{h}_\perp|^{2} + \eta},
\end{equation*}
where \(\tilde{h}_\perp\) is the Fourier transform of the extended depth PSF, \(a^{*}\) stands for the complex conjugate of \(a\), and
\(\eta\) is a regularization parameter set at \(10^{-3}\). This filter enhances high frequencies inside the OTF and removes noise beyond
the OTF cutoff. We then estimate empirically the standard deviation of these pre-filtered speckled images. Finally, an iterative,
moment-matching algorithm is used to achieve a super-resolved reconstruction of the object. We use a Tikhonov regularization method, the
reconstructed fluorescence density is given by:
\begin{equation*}
    \hat{\rho}(\boldsymbol{r}) = \arg\min_{\rho} \left( \|\hat{\sigma}(\boldsymbol{r}) - \sigma(\boldsymbol{r};\rho)\|^{2}
    + \mu \|\rho\|^{2} \right),
\end{equation*}
with \(\hat{\sigma}(\boldsymbol{r})\) the experimental standard deviation and \(\sigma(\boldsymbol{r};\rho)\) the theoretical standard
deviation derived from the expected variance. In this theoretical expression, we generate an auto-covariance function of the speckle
and PSF that matches the Fourier support of the experimental auto-covariance function of the speckle.

\subsection{Sample preparation}

\paragraph{Immunolabelling of mouse intestinal epithelium}
Sample preparation followed Mazzella et al.~\cite{mazzella2024extended}. The jejunum of a euthanized mouse was fixed in 4\% paraformaldehyde
(Electron Microscopy Sciences), washed in PBS, cut into
200--300\,\(\mu\)m thick slices and permeabilized with 1\% Triton X-100 for 1\,h. Desmoplakin immunolabelling was performed by
overnight incubation with the primary antibody (Progen, 651109, 1:50 dilution), followed by washing in PBS. The secondary antibody
(goat anti-mouse, Alexa Fluor 488, Invitrogen, A11029, 1:200 dilution) was then incubated overnight, and the sample was washed again
in PBS before mounting in Aqua-Poly/Mount (Polysciences). Mouse husbandry and procedures were performed in accordance with European
and national regulations (protocol authorization APAFIS \#38994-2023062718345638). The mice were bred in the CBI specific and
opportunistic pathogen-free (SOPF) animal facility and maintained in a C57Bl/6N genetic background.

The desmosome images analysed here were acquired using 2D-RIM with the pupil-plane phase mask described above. Desmoplakin was detected by antibody
labelling with an Alexa Fluor
488 secondary antibody.

\paragraph{Tilted prepared endothelial-cell slide}
For Fig.~\ref{fig:space_varying}, we used a permanently mounted FluoCells Prepared Slide \#2
(Invitrogen, Thermo Fisher Scientific; cat.~no.~F14781) containing bovine pulmonary artery endothelial cells (BPAECs).
In the supplied specimen, F-actin is labelled with Texas Red-X phalloidin, microtubules with a mouse anti-\(\alpha\)-tubulin
antibody and an Alexa Fluor 488-conjugated goat anti-mouse secondary antibody, and nuclei with DAPI. Only the phalloidin
channel was analysed.
The intact slide was tilted in the microscope relative to the objective focal plane, producing a defocus gradient across
the field of view.

\paragraph{Actin labelling of \textit{Drosophila} egg chambers}
\textit{Drosophila} egg chambers were fixed with 4\% paraformaldehyde (15714, Electron Microscopy Sciences) in cytoskeleton buffer prewarmed at
37\,\(^\circ\)C (10\,mM MES, pH 6.1 adjusted with NaOH, 150\,mM NaCl, 5\,mM EGTA, 5\,mM glucose, 5\,mM MgCl\(_2\)) for 15\,min and rinsed with PBS.
Cells were then permeabilized, exposed to blocking solution and incubated for 2\,h at room temperature with primary and secondary antibodies together
with 0.165\,\(\mu\)M Alexa Fluor 488-phalloidin (A12379, Thermo Fisher Scientific). Only the phalloidin channel, which labels the actin cytoskeleton,
was imaged in this work.

\subsection{Evaluation metrics}

Let \(x\) be the latent clean image, \(h_w\) the ground truth PSF and \(y\) the observed blurry image.
Our PSF estimation model produces an estimate \(\hat{h}\) of the PSF and another module produces an estimate \(\hat x\) of the sharp image from the input \(y\).
To evaluate the performance of our method, we consider two categories of metrics:
\begin{itemize}
    \item PSF estimation metrics that directly compare \(\hat{h}\) to \(h_w\),
    \item image reconstruction metrics that compare the deblurred version \(\hat{x}\) to the latent image \(x\).
\end{itemize}
In real experiments, the sample may drift between acquisitions and the exposure time or fluorescence intensity can vary.
Therefore, we always register the images and fit the best affine transformation of the grey levels before computing the metrics.

\subsubsection{PSF metrics} PSFs can be interpreted as probability distributions, since they are nonnegative and sum to one.
Hence, we can use metrics over probability spaces:
\begin{itemize}
    \item \textbf{Total variation distance:} the value \(L_1(\hat{h}, h_w) = \|\hat{h} - h_w\|_1\) yields a value in the range \([0, 2]\), which can be
        interpreted as a relative \(\ell^1\) error since the PSFs are normalized to sum to one. Note that \(\tfrac{1}{2}\|\hat{h} - h_w\|_1\) corresponds to
        the total variation distance, i.e., the fraction of mass that is displaced.
    \item \textbf{Wasserstein distances:} the Wasserstein-1 (W1) distance measures the minimum cost of transporting mass to transform one PSF into another.
        The Monge definition writes:
        \[
            \begin{gathered}
                W_1(\hat{h}, h_w) = \inf_{T} \int_{\mathbb{R}^2} \|x - T(x)\| \, h_w(x)\, dx \\
                \text{subject to} \quad T_\# h_w = \hat{h}.
            \end{gathered}
        \]
        The constraint \(T_\# h_w = \hat{h}\) means that the pushforward of \(h_w\) by the map \(T\) must equal \(\hat{h}\), ensuring that the mass is
        preserved during transportation.
        We use the POT library~\cite{flamary2021pot} to compute the Wasserstein distance. We set the pixel size to 1. Hence, for two Dirac masses separated
        by \(d\) pixels, the Wasserstein-1 distance is equal to \(d\).
    \item \textbf{Reblurring metrics:} if the latent image \(x\) does not have enough frequency content (e.g. constant image), it is impossible to
        recover the PSF.
        Therefore, it is instructive to evaluate the recovery dependence in accordance with the frequency content of the latent image.
        To this end, we also evaluate the PSNR, SSIM or LPIPS metrics between the convolution of the latent image with the estimated PSF and the convolution
        with the ground truth PSF.
        Letting \(M\) be one of these metrics, we define the reblurring metric as
        \begin{equation}
            M_{\mathrm{blur}}(\hat{h}, h_w) = M(x \star h_w, x \star \hat{h}).
        \end{equation}
\end{itemize}

\subsubsection{End-task metrics}
While accurate PSF estimation is important, the ultimate goal in computational imaging is to improve the quality of the reconstructed image.
Hence, we also report the image quality metrics \(M(\hat x, x)\), where \(M\) is one of the standard image quality metrics such as PSNR or SSIM, and
\(\hat x\) is the deconvolved image obtained by applying a non-blind deconvolution algorithm using the estimated PSF \(\hat h\).
This allows us to assess how errors in PSF estimation translate into reconstruction performance.

\newpage
\subsection{Segmentation evaluation}

We evaluated the effect of restoration on segmentation using the full test set of the Cellpose
dataset~\cite{stringer2021cellpose} (\(n = 68\) images), which was used to train neither \name nor
the Cellpose segmentation models in Extended Data Fig.~\ref{edfig:segmentation}.

Each image was
degraded by a PSF generated with our microscope simulation model (see Section~\ref{sec:image_formation}).
Degraded images were restored either with \name{}-RAM, or with the Cellpose3 restoration
network~\cite{stringer2025cellpose3}. We segmented the
degraded, restored and original sharp images with CellSAM, Cellpose-SAM,
and the cyto3, cyto2 and cyto models, all with default parameters.

Segmentation accuracy was quantified by the average precision (AP) as defined in
Cellpose~\cite{stringer2021cellpose}: at a given intersection-over-union (IoU) threshold, predicted
and ground-truth masks are matched one-to-one, and \(\mathrm{AP} = \mathrm{TP}/(\mathrm{TP}+\mathrm{FP}+\mathrm{FN})\).
AP was computed per image and averaged over the test set for IoU thresholds from 0.5 to 0.95.
The dependence on noise was evaluated for Cellpose-SAM at an IoU threshold of 0.5, with the gain
varied from \(2^8\) to \(2^{14}\). For each segmentation model, we compare AP across inputs to assess the benefit of restoration;
the experiment is not intended to rank the segmentation models against one another.
Restoration with \name{}-RAM increased the average precision for all models.

\subsection{BioSR restoration evaluation}

BioSR~\cite{qiao2021biosr} is a public dataset designed to benchmark deep-learning super-resolution in fluorescence
microscopy. For each field of view, it provides a diffraction-limited widefield image, obtained by summing the raw SIM
frames, together with the corresponding SIM reconstruction. It covers clathrin-coated pits, endoplasmic reticulum,
microtubules and F-actin (the latter also imaged with nonlinear SIM), each at nine signal levels. We used it to evaluate
restoration on real data acquired independently of our instruments, for which a higher-resolution image of the same
field is available.

The widefield images are not deliberately aberrated: their blur is that of the actual microscope PSF, which is not
provided. The task is therefore blind deconvolution. For each widefield image, \name{} estimated the PSF and RAM
restored the image using this estimate. The SIM reference was downsampled to the widefield pixel grid, and PSNR, SSIM
and LPIPS were computed against this reference for both the widefield input and its restoration; the reported gain is
the difference between the two scores. To assess robustness to noise, we also applied \name{} to the same field at the
nine signal levels (Extended Data Fig.~\ref{edfig:biosr_gains}b). Neither \name{} nor RAM was trained or fine-tuned on
BioSR.

This evaluation has two limitations. BioSR provides no reference PSF, so it measures restoration quality rather than
PSF accuracy. In addition, the SIM reference contains spatial frequencies beyond the widefield cutoff, which
deconvolution cannot recover; absolute scores therefore remain well below a perfect match, and only the gains are
informative.

\subsection{Zernike-range generalization}

We compared six models trained with Zernike ranges labelled \(3\)--\(6\), \(3\)--\(10\), \(3\)--\(15\), \(3\)--\(21\), \(3\)--\(28\) and
\(3\)--\(36\), and evaluated each model on all six aberration ranges at Poisson gain \(\gamma=2^{15}\) (Extended Data
Fig.~\ref{edfig:zernike_generalization}). The displayed statistics are medians. In addition to PSF-estimation errors, we compared images reblurred
with the estimated and true PSFs using the reblurring metrics defined above.

\subsection{Quantitative analysis of the tilted specimen}

A nine-plane \(z\)-stack of the field was acquired; each plane is the sum of 200 RIM frames (97\,nm pixels). On tiles of
\(100 \times 64\) pixels, image sharpness was measured by the variance of a Laplacian-of-Gaussian filter (\(\sigma = 1.5\) pixels)
normalized by the squared mean intensity. The best-focus plane of each tile was obtained by a Gaussian fit of the sharpness around its
maximum across planes, and a plane was fitted to these positions over the field (residual 0.32 planes RMS). \name{} was applied to
each plane with the following tiling: \(128 \times 128\)-pixel patches with a stride of 64 pixels, i.e.\ a \(12 \times 12\)
grid on the \(800 \times 800\)-pixel field; the last row and column, which extend into the padded image border, were excluded,
leaving 121 patches per plane and 1{,}089 in total. The defocus \(\Delta z\) of each patch (Fig.~\ref{fig:space_varying}b) is the difference between
the fitted plane at the patch centre and the plane index, multiplied by the calibrated plane
spacing. PSF width
is the diameter of the smallest circle enclosing 50\% of the PSF intensity, over all sub-pixel positions of its centre (EE50),
computed on the PSF upsampled four times.
A Spearman rank correlation of \(\rho = 0.94\) was computed between the EE50 of all patches and their absolute defocus. The defocus model (dashed line
in Fig.~\ref{fig:space_varying}b) is the EE50 of an aberration-free scalar PSF (NA~1.49, \(\lambda = 615\)\,nm, immersion index 1.518) with defocus
\(\Delta z\), sampled like the estimates (\(71 \times 71\) pixels of 97\,nm); in focus, it is 238\,nm.

The plane spacing, nominally 250\,nm, was calibrated independently of \name{}. For each tile, the width (full width at half maximum
of a fitted Gaussian) of the measured sharpness curve, in planes, was compared with that of the same sharpness metric computed on a
simulated defocus series, the in-focus RIM image of the tile convolved with the defocused scalar PSF, in nm. The median ratio over the
95 tiles gives 172\,nm per plane (interquartile range 162--183\,nm). The band around the defocus model shows the 16th--84th
percentiles of the model EE50 when the defocus of a patch is drawn from a Gaussian centred on \(\Delta z\), of standard deviation
\(\sqrt{\sigma_0^2 + (\epsilon \Delta z)^2}\): \(\sigma_0 = 0.32\) planes (56\,nm) is the residual of the plane fitted to the best-focus
positions, and \(\epsilon = 6\%\) the half interquartile range of the calibration relative to its median. The dotted lines mark the
defocus range of the training PSFs for the defocus mode alone: training PSFs have a Zernike defocus coefficient of at most 0.15 waves
RMS, which corresponds to \(|\Delta z| = 287\)\,nm for this objective (projection of the high-NA defocus wavefront on Zernike defocus
and primary spherical aberration).

\subsection{Intensity profiles of RIM reconstructions}

Profiles in {Extended Data Fig.~\ref{edfig:rim_egg_chamber}} were sampled perpendicular to the actin cables, whose orientation was obtained from the
structure tensor
of each detail, over 0.95\,\(\mu\)m (12 pixels of 79\,nm), with bilinear interpolation every half pixel and averaging over three
parallel lines one pixel apart; each profile is normalized to its maximum, such that the dip strength is defined as the intensity of the mimimum.
Lines were placed across closely spaced cables. The separation of the two components is the distance between the peaks, given the
0.04\,\(\mu\)m sampling of the profiles.

\subsection{Plane-by-plane restoration of a zebrafish stack}

A zebrafish volume acquired with spinning-disk confocal microscopy was processed independently in each \(xy\) plane. For each plane, \name
estimated the lateral PSF and RAM performed two-dimensional deconvolution. The restored planes were assembled in their original axial order, and
corresponding \(xy\), \(xz\) and \(yz\) sections were displayed for qualitative comparison (Extended Data Fig.~\ref{edfig:zebrafish}).

\subsection{Use of large language models}

During the preparation of this manuscript, the authors used a large language model to assist with structuring and drafting the text. The authors
reviewed and edited all content and take full responsibility for the scientific content, the analyses and the conclusions of this work.

\clearpage
\onecolumn
\section*{Extended Data}
\setcounter{figure}{0}
\renewcommand{\thefigure}{\arabic{figure}}
\renewcommand{\theHfigure}{extended.\arabic{figure}}
\captionsetup[figure]{name=Extended Data Fig.}
\setcounter{table}{0}
\renewcommand{\theHtable}{extended.\arabic{table}}
\captionsetup[table]{name=Extended Data Table}

\begin{figure}[!ht]
    \centering
    \includegraphics[width=\textwidth]{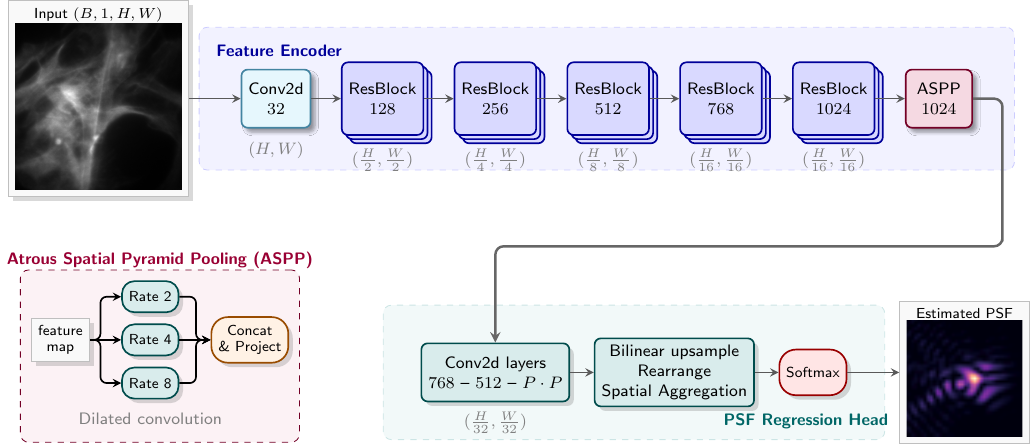}
    \caption{\textbf{Architecture of \name.} \name is a fully convolutional residual network designed for robust PSF estimation. The model takes as
        input a blurry image
        patch of size \(H \times W\) with one to three channels and outputs one estimated PSF of size \(P \times P\) per channel. It consists of a
        feature encoder with \(5\) residual blocks, a
        multi-scale context aggregation module, and a PSF regression head. The number of feature channels at each stage is indicated at the bottom of each
    block. The model comprises approximately \(98\) million parameters.}
    \label{edfig:architecture}
\end{figure}

\clearpage

\begin{figure}[!ht]
    \centering
    \includegraphics[width=0.95\textwidth]{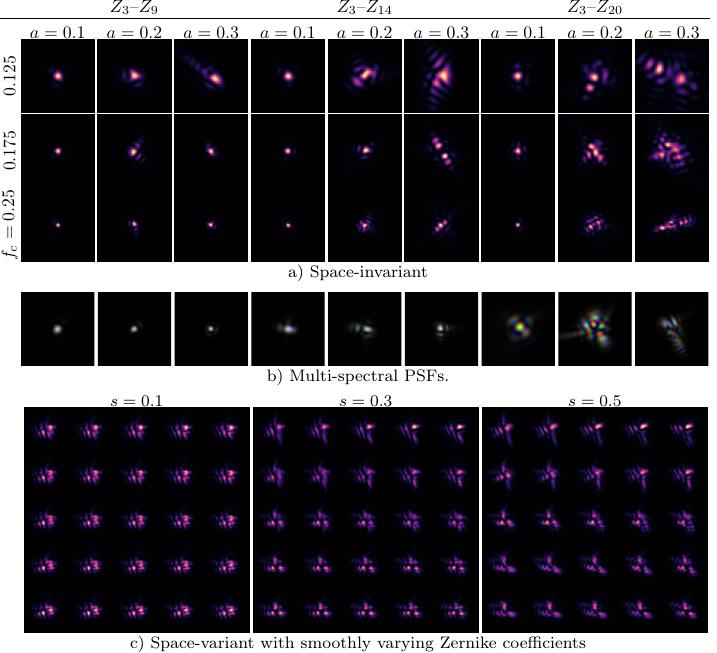}
    \caption{\textbf{Simulated PSFs.} Illustration of the parameter effects in our PSF simulations. (\textbf{a})~Varying the cutoff frequency \(f_c\), the
        amplitude \(a\) and the highest Zernike mode. (\textbf{b})~Three-channel PSFs with \(\lambda_{\textrm{red}} = 650\)\,nm,
        \(\lambda_{\textrm{green}} = 550\)\,nm and \(\lambda_{\textrm{blue}} = 450\)\,nm for \(Z_3\)--\(Z_{35}\), at various cutoff frequencies
        and amplitudes. (\textbf{c})~Spatially varying PSFs for increasing strength \(s\) of the spatial variations (not used to train the
    estimator).}
    \label{edfig:example_PSFs}
\end{figure}

\clearpage

\begin{figure}[!ht]
    \centering
    \begin{tikzpicture}
        \node[anchor=south west,inner sep=0] (heatmaps) at (0,0)
        {\includegraphics[width=\textwidth]{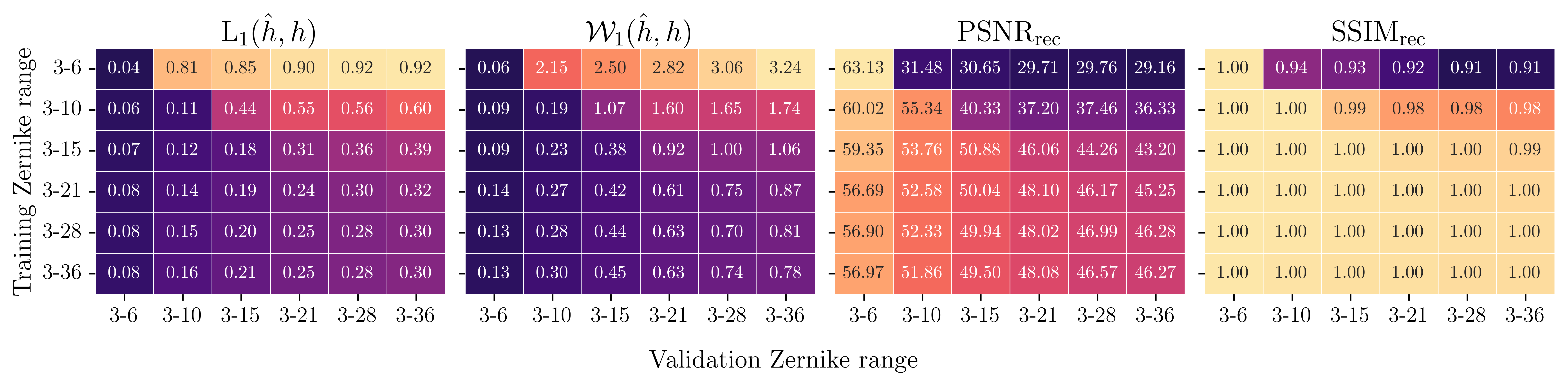}};
        \begin{scope}[x={(heatmaps.south east)},y={(heatmaps.north west)}]
            \fill[white] (0.52,0.888) rectangle (1,1);
            \node[font=\fontsize{8}{9}\selectfont] at (0.642,0.920) {\(\mathrm{PSNR}_{\mathrm{blur}}\)};
            \node[font=\fontsize{8}{9}\selectfont] at (0.884,0.920) {\(\mathrm{SSIM}_{\mathrm{blur}}\)};
        \end{scope}
    \end{tikzpicture}
    \caption{\textbf{Generalization across training and evaluation Zernike ranges.} Rows indicate the Zernike range used to train each model;
        columns indicate the range used for evaluation. Entries are medians at Poisson gain \(\gamma=2^{15}\). From left to right: \(L_1\) PSF error,
        Wasserstein-1 PSF error, reblurring PSNR and reblurring SSIM. The latter two metrics compare images convolved with the estimated and true PSFs,
        rather than deconvolved images. Training on a restricted range leads to substantially larger errors on higher-order aberrations. Broader training
    improves robustness across the tested ranges while retaining high reblurring fidelity on simpler aberrations.}
    \label{edfig:zernike_generalization}
\end{figure}

\clearpage

\begin{figure}[!ht]
    \centering
    \FigureWithScaleBars[\textwidth]{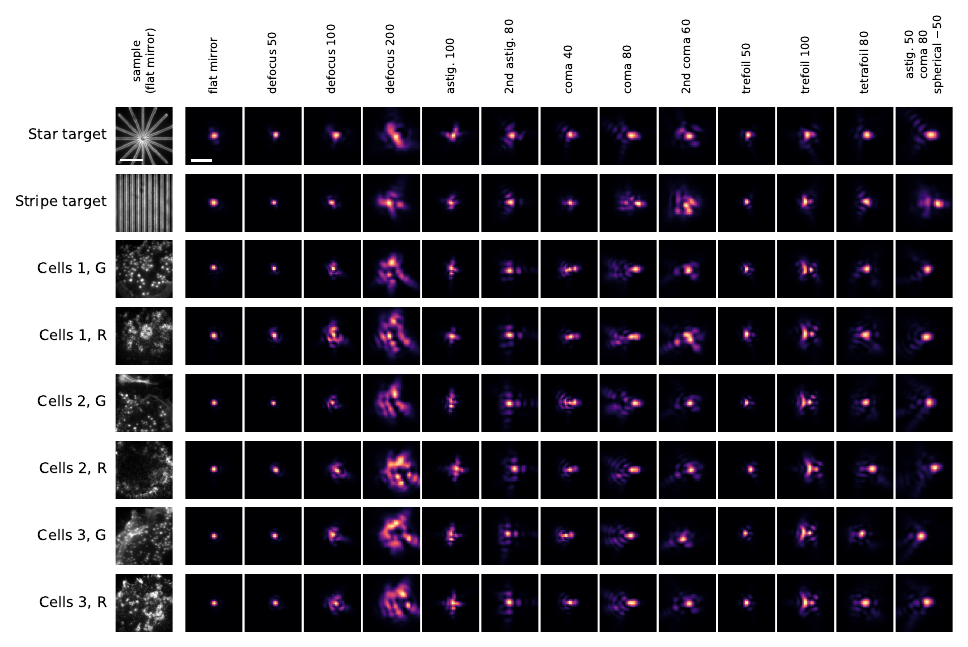}{1400}{936}{\ReferenceMosaicScaleLabels}
    \caption{\textbf{Reference PSFs of the deformable-mirror dataset.} Reference PSFs estimated from paired flat-mirror and
        aberrated acquisitions, for one field of view of each sample (rows; G and R: green and red channels) and 13 of the 32 mirror
        states (columns). The left column shows the flat-mirror image of each field of view, from which, together with the
        corresponding aberrated images, the reference PSFs are estimated. Column labels give the Zernike modes set on the mirror (2nd: secondary
        aberration) and their amplitudes, in the
        units of the mirror control software. PSFs in each column share the same commanded mirror setting. Their shapes show
        common features across samples together with acquisition-to-acquisition variation; repeated commands do not guarantee
        identical realised aberrations. The comparison motivates estimating a reference PSF separately for each acquisition. Each PSF is
        normalized to its maximum; the central
    \(41 \times 41\) pixels (\(2.7\,\mu\)m) are shown. Pixel size, 65\,nm for both the images and the PSFs.}
    \label{edfig:estimated_PSFs_DM}
\end{figure}

\clearpage

\begin{figure}[!ht]
    \centering
    \includegraphics[width=\textwidth]{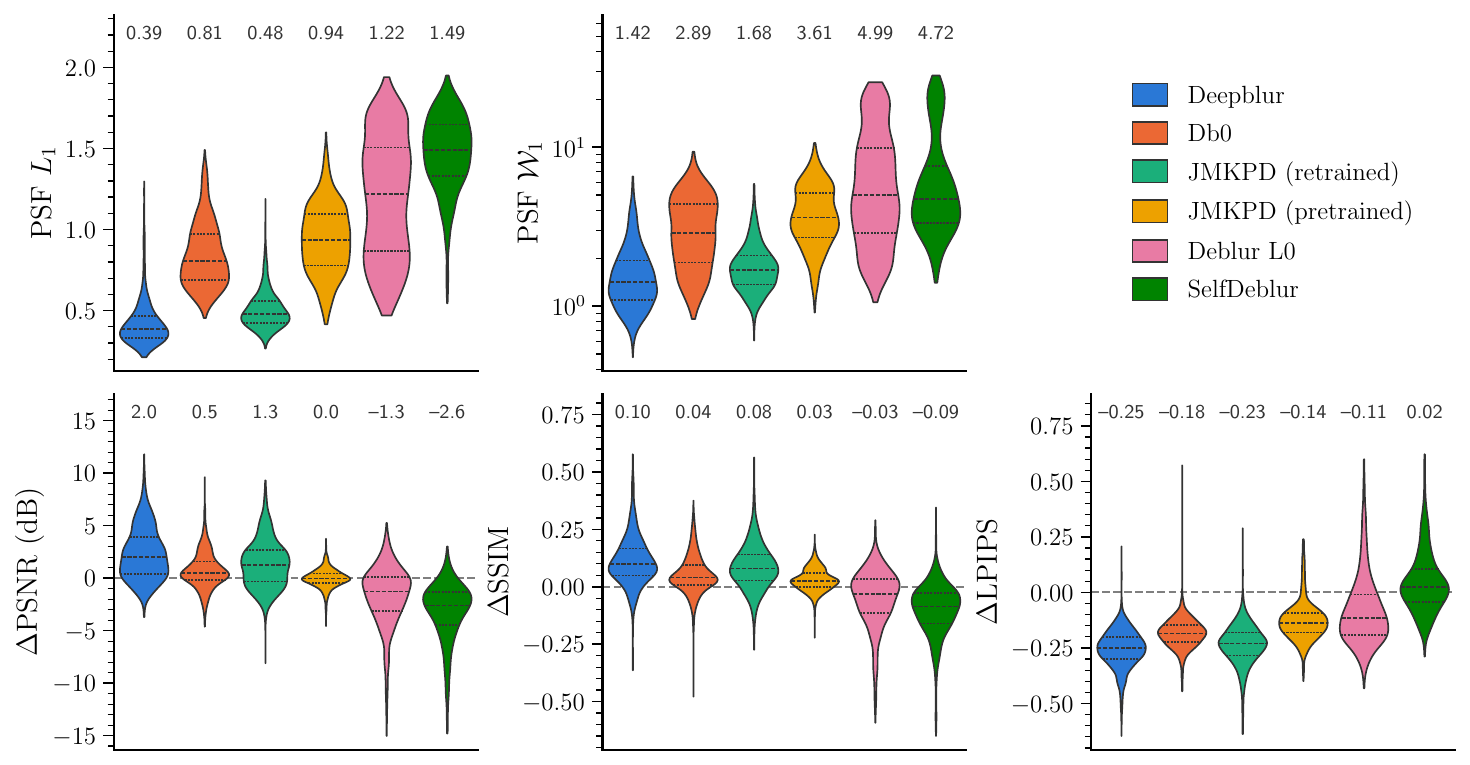}
    \caption{{\textbf{PSF-estimation errors and deconvolution gains on the deformable mirror dataset.} (\textbf{Top})~\(L_1\) distance
            (left) and Wasserstein-1 distance \(W_1\) in pixels (middle, logarithmic scale) between the PSF estimated by each method and the reference
            PSF; lower is better. (\textbf{Bottom})~Change \(\Delta M = M(\hat{x}, x) - M(y, x)\) in PSNR (left), SSIM (middle) and LPIPS (right),
            where \(y\) is the aberrated image, \(\hat{x}\) its deconvolution by the same fine-tuned RAM model~\cite{terris2026reconstruct} driven by
            each method's PSF, and \(x\) the reference sharp image (Fig.~\ref{fig:dm}b). Positive \(\Delta\)PSNR and \(\Delta\)SSIM and negative
            \(\Delta\)LPIPS indicate an improvement over the aberrated input; the dashed horizontal line marks no change. Methods, in the same order in
            the legend and in every panel: \name, Db0~\cite{debarnot2024deep}, JMKPD~\cite{carbajal2023blind} retrained with our training procedure and
            with its original pretrained weights, L0~\cite{pan2014deblurring} and SelfDeblur~\cite{ren2020neural}. Statistics over the 1{,}290 aberrated
            images; the 42 acquisitions with a flat mirror are excluded. In each violin, the dashed line marks the median and the dotted lines the first
    and third quartiles; the number above each violin is its median.}}
    \label{edfig:dm_violins}
\end{figure}

\clearpage

\begin{figure}[!ht]
    \centering
    \includegraphics[width=\textwidth]{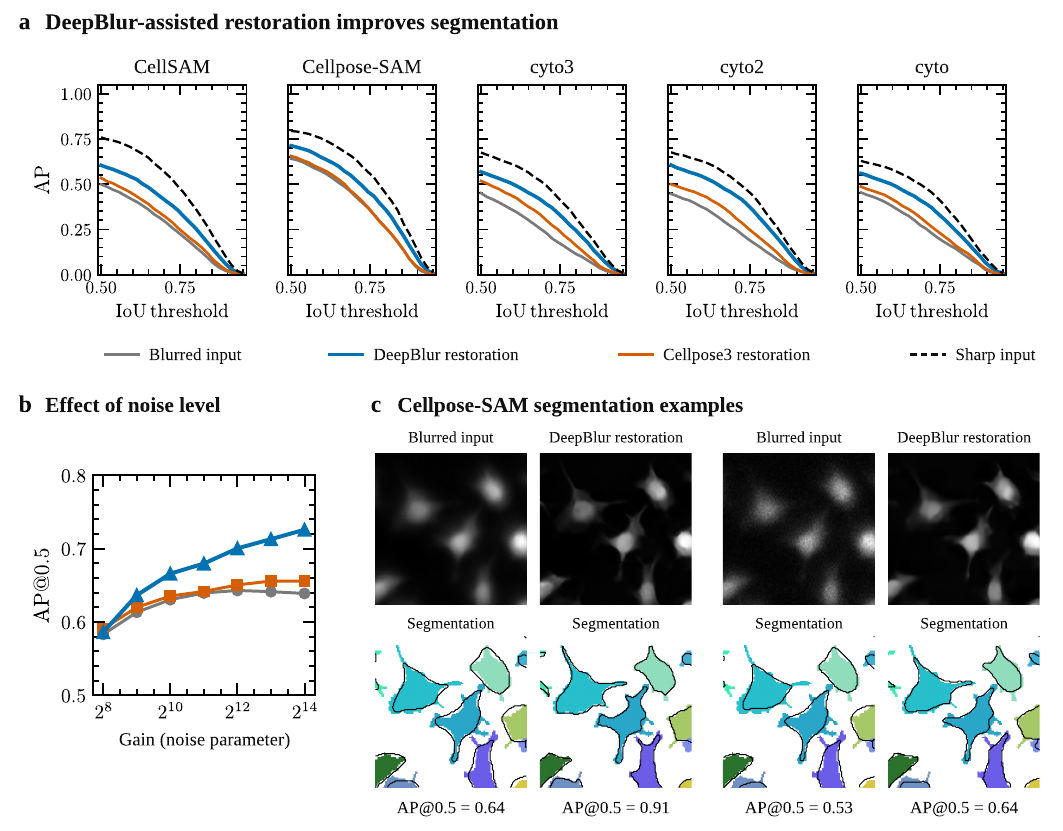}
    \caption{\textbf{\name{}-RAM restoration improves segmentation of synthetically degraded images.}
        (\textbf{a})~Segmentation average precision (AP) as a function of the intersection-over-union (IoU) threshold
        for CellSAM, Cellpose-SAM, cyto3, cyto2 and cyto on the Cellpose test set (\(n = 68\) images; gain \(=2^{11}\)).
        Inputs are the degraded images (grey), images restored by RAM using \name-estimated PSFs (blue), images restored
        with Cellpose3~\cite{stringer2025cellpose3} (orange), and the original sharp images (black dashed).
        (\textbf{b})~Cellpose-SAM AP at an IoU threshold of 0.5 as a function of the gain controlling the simulated
        noise level (higher gain, lower noise); colours as in (\textbf{a}).
        (\textbf{c})~Degraded images and \name{}-RAM restorations of the same field at gains \(2^{11}\) (left) and \(2^8\) (right),
        with the corresponding Cellpose-SAM segmentations (bottom; filled masks, predictions; black outlines, ground truth).
    AP at an IoU threshold of 0.5 is given below each segmentation.}
    \label{edfig:segmentation}
\end{figure}

\clearpage

\begin{figure}[!ht]
    \centering
    \makebox[\textwidth][l]{\textbf{(a)}}\\
    \includegraphics[width=\textwidth]{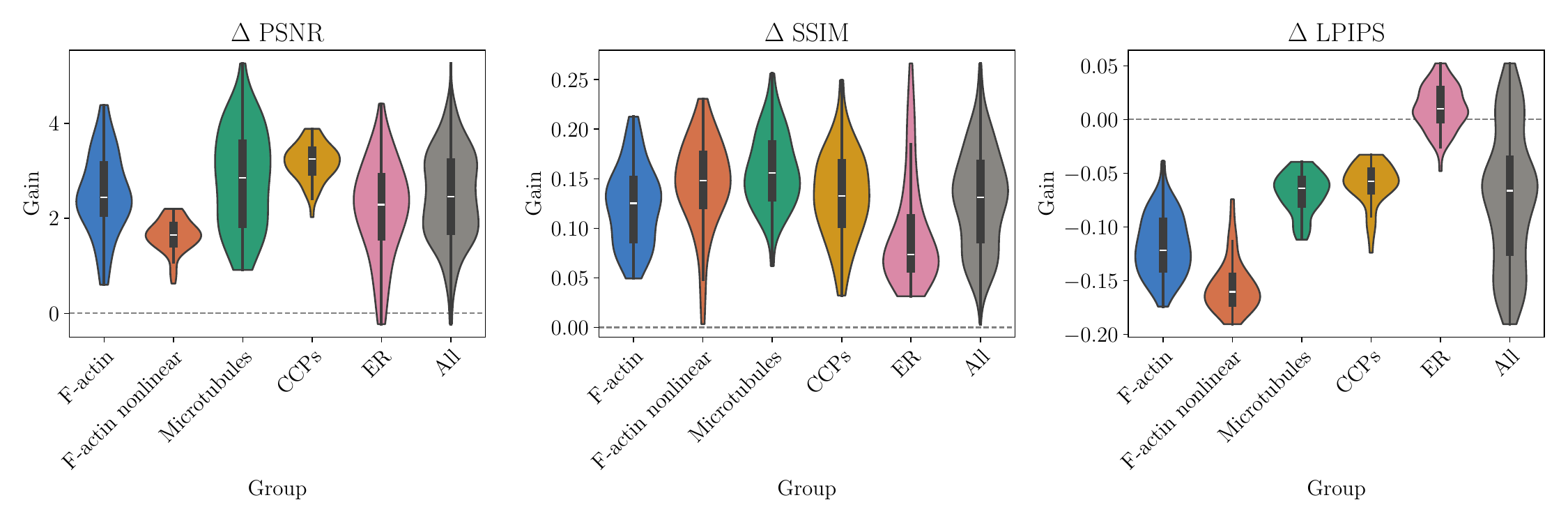}\\
    \makebox[\textwidth][l]{\textbf{(b)}}\\
    \includegraphics[width=\textwidth]{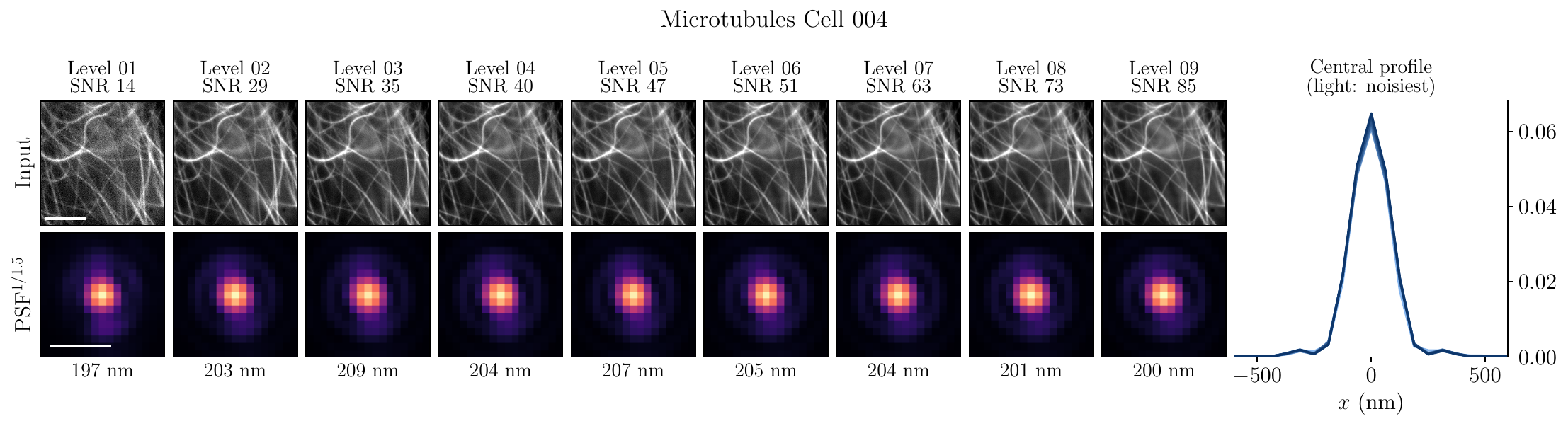}
    \caption{\textbf{Restoration gains and robustness of PSF estimation to noise on the BioSR dataset.}
        (\textbf{a})~Restoration gains. For each low-resolution BioSR image~\cite{qiao2021biosr}, PSNR, SSIM and LPIPS
        were evaluated against its paired super-resolved reference after downsampling the reference to the input pixel grid. The plotted gain for each
        metric \(M\) is \(\Delta M=M(\hat{x},x_{\mathrm{ref}})-M(y,x_{\mathrm{ref}})\), where \(y\) is the low-resolution input, \(\hat{x}\) is its
        restoration by RAM using the PSF estimated by \name, and \(x_{\mathrm{ref}}\) is the downsampled super-resolved reference. Positive gains in PSNR
        and SSIM and negative gains in LPIPS indicate improvement. Results are grouped by specimen type and imaging condition: F-actin, nonlinear-SIM
        F-actin, microtubules, clathrin-coated pits (CCPs) and endoplasmic reticulum (ER); All pools these groups. Each violin shows the distribution of
        gains, its inner box the median and quartiles, and the dashed line no change.
        LPIPS (Learned Perceptual Image Patch Similarity), a perceptual metric developed for natural images, is shown as a complementary measure;
        median PSNR and SSIM improved in all groups. This comparison assesses restoration against a higher-resolution
        reference, not the accuracy of the estimated PSF.
        (\textbf{b})~Robustness of the PSF estimation to noise, for one microtubule field (Cell 004) acquired at the nine BioSR noise levels.
        Top, the \(256\times256\)-pixel crop given to \name{} (sum of the nine raw SIM frames; the same crop at every level), with its signal-to-noise
        ratio (SNR) above. Bottom, the single PSF that \name{} estimates from that crop, shown with a gamma correction of \(1/1.5\), with its full
        width at half maximum (FWHM, from a 2D Gaussian fit) below. Right, central horizontal profiles of the nine
    PSFs on a linear scale (lighter, noisier).}
    \label{edfig:biosr_gains}
\end{figure}

\clearpage

\begin{figure}[!ht]
    \centering
    \begin{tikzpicture}
        \node[anchor=south west,inner sep=0,outer sep=0] (scaleimage) at (0,0)
        {\includegraphics[width=\textwidth,trim=0 350bp 0 -2bp,clip]{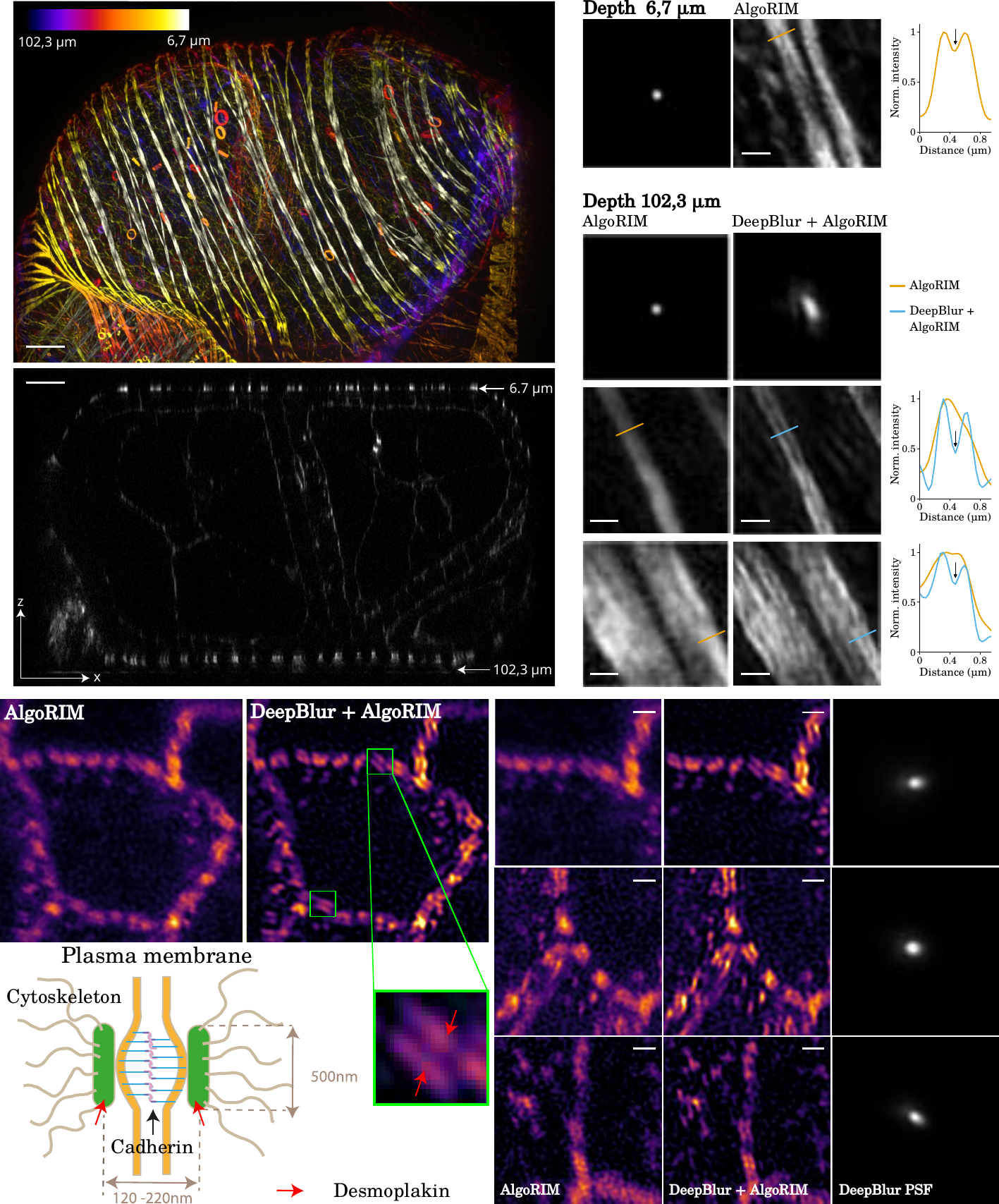}};
        \begin{scope}[x={(scaleimage.south east)},y={(scaleimage.north west)}]
            \def\ScaleCanvasWidth{681}%
            \def\ScaleCanvasHeight{472}%
            \EggChamberScaleLabels
        \end{scope}
    \end{tikzpicture}
    \caption{{\textbf{Improved RIM resolution in a Drosophila egg chamber using \name-estimated PSFs.}
            The actin-rich muscle sheath of a fixed stage-9 \textit{Drosophila} egg chamber was imaged near the coverslip and at depth.
            Left, depth-coloured overview and \(x\)-\(z\) view locating the two regions. Right, local PSFs, image details and intensity profiles.
            Near the coverslip, at \(6.7\,\mu\mathrm{m}\), AlgoRIM with the theoretical PSF separates adjacent actin cables.
            At \(102.3\,\mu\mathrm{m}\), the theoretical PSF is compared with local PSFs estimated by \name{} from the same acquisitions.
            The broader, asymmetric estimated PSFs improve cable separation, revealing an organisation resembling that near the coverslip.
            The near and far sides provide an internal structural comparison, not paired ground-truth images.
            Normalized profiles along the coloured lines compare AlgoRIM (orange) and \name{} + AlgoRIM (blue).
            Near the coverslip, two cables 0.28\,\(\mu\)m apart are separated by a 0.18 dip. At depth, AlgoRIM with the theoretical PSF
            gives a single peak or a flat top, whereas \name{} + AlgoRIM separates components 0.32 and 0.28\,\(\mu\)m apart,
    with dips of 0.47 and 0.22, respectively.}}
    \label{edfig:rim_egg_chamber}
\end{figure}

\clearpage

\begin{figure}[!ht]
    \centering
    \includegraphics[width=\textwidth,trim=0 102bp 0 0,clip]{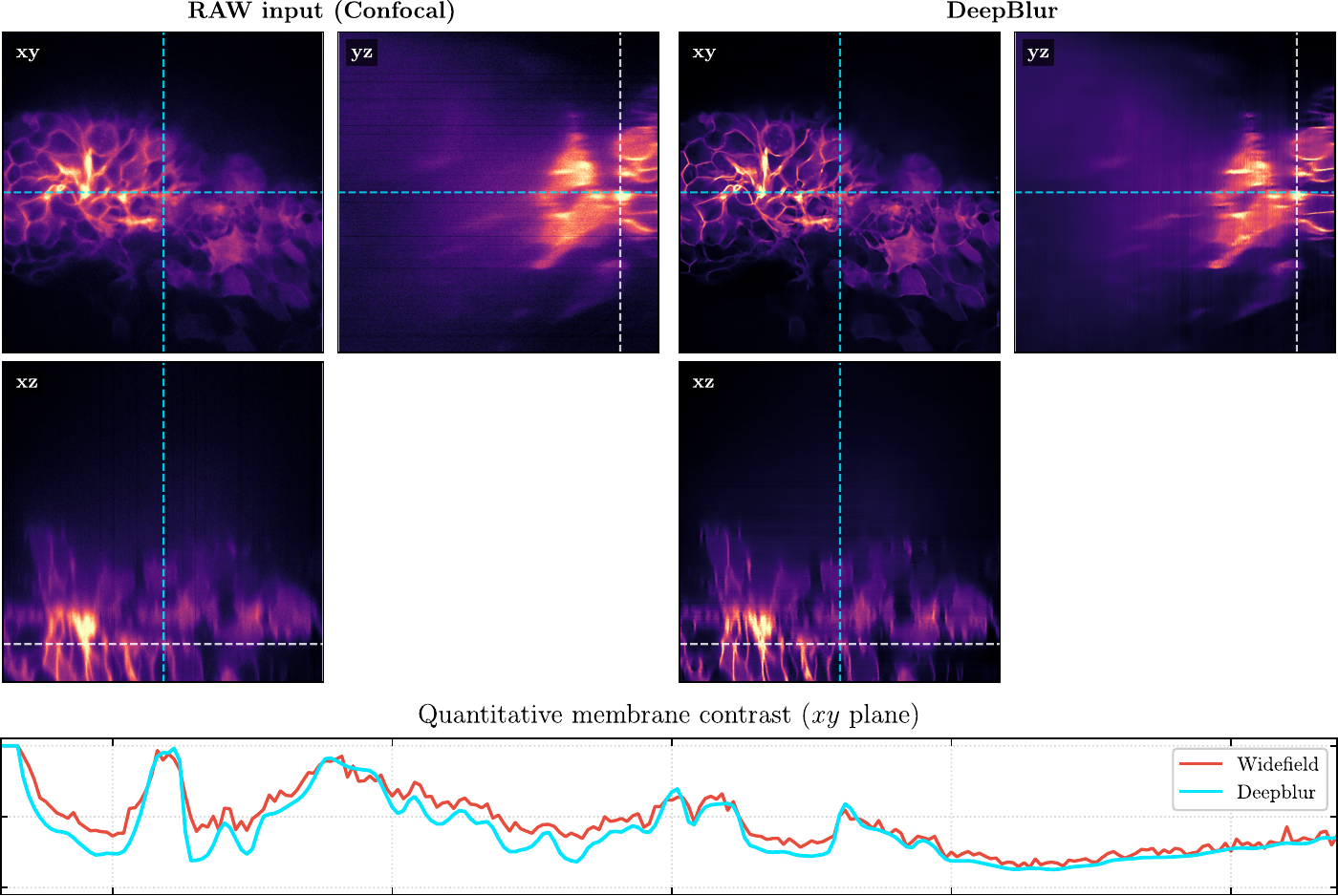}
    \caption{\textbf{Plane-by-plane lateral restoration of a zebrafish volume acquired with spinning-disk confocal microscopy.}
        Corresponding \(xy\), \(yz\) and \(xz\) sections of the raw acquisition (left) and the volume after independent two-dimensional PSF estimation
        with \name and deconvolution with RAM in each \(xy\) plane (right). Dashed lines mark the locations of the displayed sections. Membrane
        structures are more distinct in the restored lateral image. Orthogonal sections show the appearance of the assembled stack; no three-dimensional
    PSF estimation or axial deconvolution is performed.}
    \label{edfig:zebrafish}
\end{figure}

\begin{table}[!ht]
    \centering
    \caption{\textbf{Parameter distributions for the simulations at training time.} The cutoff frequency \(f_c\) refers to the shortest
    wavelength.}\label{tab:simulation_parameters}
    \footnotesize
    \setlength{\tabcolsep}{4pt}
    \begin{tabular}{l *{8}{c}}
        \toprule
        & Bit depth       & Poisson gain          & Gauss. std            & Zernike modes           & Zernike scale                     & Cutoff freq. &
        Wavelength          & Chromatic pert.
        \\
        \midrule
        Notation            & \(\mathrm{b}\)  & \(\gamma\)            & \(\sigma\)            & \(K\)                   & \(a\)
        & \(f_c\)
        & \(\lambda_c\)   & \(\varepsilon_{k,c}\)                                                                                                        \\
        Value               & 16              & \(2^5\)--\(2^{12}\)   & \(0.5\)--\(5\) counts & \(n_k = 2\)--\(7\) (33) & \(0.1\)--\(0.3\)                  &
        \(0.125\)--\(0.25\) & 400--700\,nm    & std \(0.005\)
        \\
        Distribution        & fixed           & log-unif.             & log-unif.             & fixed                   &
        See~\eqref{eq:zernike_coeff_dist} & unif.
        & unif.           & Gaussian                                                                                                                     \\
        \bottomrule
    \end{tabular}
\end{table}

\end{document}